\documentclass[namedreferences,hyperref,optionalrh,solaromanenum]{spr-sola}
\usepackage{amsmath}
\usepackage{graphicx}  

\usepackage{amssymb}                    
\usepackage{color}  
\usepackage{xcolor,colortbl}
\definecolor{lavender}{rgb}{0.9, 0.9, 0.98}
\usepackage{pgf}

\chardef\us=`\_

\begin{document}

\begin{frontmatter}

\title{An efficient method for magnetic field extrapolation based on a family of analytical three-dimensional magnetohydrostatic equilibria}

%

\author[addressref=aff1,email={lmn6@st-andrews.ac.uk}]{\inits{L.N.}\fnm{Lilli}~\snm{Nadol}\orcid{0000-0003-0972-3865}}
\author[addressref=aff1,email={tn3@st-andrews.ac.uk}]{\inits{T.N.}\fnm{Thomas}~\snm{Neukirch}\orcid{0000-0002-7597-4980}}

%
\runningauthor{L. Nadol, T. Neukirch}
\runningtitle{Efficient MHS extrapolation}

\address[id=aff1]{School of Mathematics and Statistics, University of St Andrews KY16 9SS, United Kingdom}

\begin{abstract}
With current observational methods it is not possible to directly measure the magnetic field in the solar corona with sufficient accuracy. Therefore, coronal magnetic field models have to rely on extrapolation methods using photospheric magnetograms as boundary conditions. In recent years, due to the increased resolution of observations and the need to resolve non-force-free lower regions of the solar atmosphere, there have been increased efforts to use magnetohydrostatic (MHS) field models instead of force-free extrapolation methods. Although numerical methods to calculate MHS solutions can deal with non-linear problems and hence provide more accurate models, analytical three-dimensional MHS equilibria can also be used as a numerically relatively “cheap” complementary method. In this paper, we present an extrapolation method based on a family of analytical MHS equilibria that allows for a transition from a non-force-free region to a force-free region. We 
demonstrate how asymptotic forms of the solutions can help to increase the numerical efficiency of the method. 
Through both artificial boundary condition testing and a first application to observational data, we validate the method's effectiveness and practical utility.
\end{abstract}


\end{frontmatter}

\section{Introduction}

%
Magnetic field extrapolation techniques are important tools for solar physics. This is due to the fact that,
despite some progress 
\citep[e.g.][]{Landi:etal2020},
it is still not possible to obtain direct measurements of the coronal magnetic field 
routinely with sufficient resolution and accuracy. Instead, extrapolation uses 
magnetohydrodynamic (MHD) equilibrium solutions together with boundary
conditions provided by photospheric magnetogram data to determine the magnetic field in the solar corona
\citep[e.g.][]{Wiegelmann2017,WiegelmannSakurai2021}.

Because the (lower) corona usually has plasma beta ($\beta_P$) values that are (significantly) below unity, traditionally 
extrapolation methods have been based on force-free magnetic fields \citep[e.g][]{Wiegelmann2017,WiegelmannSakurai2021}.
However, with the resolution of the data available now it might be necessary for extrapolation methods to take into account
the lower layers of the solar atmosphere, which are not force-free \citep[e.g.][]{Metcalf1995}. 
%
%
Therefore, in recent years magnetohydrostatic (MHS) extrapolation methods have been used as an
alternative to 
force-free models,
because they
can accommodate for a non-force-free photosphere and chromosphere
\citep[for an overview, see][]{Zhu:etal2022}.

The MHS equations are given by
\begin{align*}
\textbf{j} \times \textbf{B} - \nabla p - \rho \nabla \Psi &= 0, \\
\nabla \times \textbf{B} &= \mu_0 \textbf{j}, \\
\nabla \cdot \textbf{B} &= 0,
\end{align*}
where $\mathbf{B}$ denotes the magnetic field, 
$\mathbf{j}$ the current density, $p$ the plasma pressure, $\rho$ the mass density, 
$\Psi$ the gravitational potential, and $\mu_0$ the permeability of free space. 
In this paper, we will use a Cartesian coordinate
system with $z$ being the height above the solar surface. 
With a constant gravitational acceleration, $g$, acting 
in the negative $z$-direction, the gravitational potential takes the form
$\Psi=gz$.

In general, these equations are non-linear and therefore
numerical
methods 
are required
to solve 
them
for extrapolation purposes
\citep[e.g.][]{Gilchrist2013, Zhu2018}, which is 
computationally expensive.
In this paper we shall discuss the alternative possibility of using analytical 3D MHS solutions 
for extrapolation purposes. 
To be able to solve the 3D MHS equations analytically one has to make a number 
of simplifying assumptions, which lead to a linear mathematical problem not only allowing for analytical
solutions but also for the superposition of different solutions. This usually allows for a quicker and
more efficient way of MHS extrapolation, but the assumptions made to simplify the mathematical problem
may also limit the applicability of extrapolation methods based on analytical MHS solutions. Such methods
should therefore not be viewed as a replacement of numerical methods based on the full non-linear MHS problem, but as complementary to those methods.


The general method we use to calculate analytical 3D MHS solutions is based on a series of papers by \citet{Low1985,Low1991,Low1992} \citep[see also][]{BogdanLow1986,Low1993a,Low1993b,Low2005}. Magnetohydrostatic equilibria determined by applying this general method in Cartesian, cylindrical and spherical coordinate systems have been used by 
many
authors
\citep[e.g.][]{Bagenal1991,Gibson1995,Gibson1996,Neukirch1995,
Neukirch1997,Neukirch2009,Aulanier1998,Aulanier1999,
Zhao1993,Zhao1994,Zhao2000,Petrie2000,Ruan2008,
AlSalti2010, AlSalti2010b, Gent2013,MacTaggart2016,Wilson2018}.

\cite{Low1991, Low1992} introduced a specific class of analytical MHS magnetic fields 
for which the non-force-free terms decay exponentially 
with increasing height.
Such a transition 
generally agrees
with the change of the plasma beta, $\beta_P$, in the solar atmosphere \citep[e.g.][]{Gary2001}. 
These solutions have been
used for 
magnetic field extrapolation using Sunrise/IMaX
data as boundary conditions 
\citep{Wiegelmann2015,Wiegelmann2017b}. 

More recently, \cite{Neukirch2019} have proposed 
a new family of MHS solutions 
showing a transition from a non-force-free to
a force-free domain.
This transition is based on a hyperbolic tangent dependence on height $z$ rather than an exponential dependence as in \cite{Low1991, Low1992}.
The family of solutions by \cite{Neukirch2019} allows
more control over the properties of the transition. However, this increase in control is gained at the
expense of an increase in model parameters.
In particular, the \cite{Neukirch2019} solutions allow for the height at which the non-force-free to force-free transition occurs and the width over which it takes place to be specified. Hence, it is possible to fit 
the transition in the nature of the magnetic field to the 
height of the transition region between
the chromosphere and the solar corona.

So far the MHS solutions by \citet{Neukirch2019} have only been applied to a simple bipolar "toy" magnetogram
and not to real data.
The main reason for this is that 
the analytical solution involves 
hypergeometric functions and these can be difficult
to evaluate numerically for 
certain parameter combinations. The main aim of this
paper is to resolve these problems and to apply the
new solution family to observational data for
the first time. 
To achieve this aim, we have found a method to 
approximate the hypergeometric functions by a piecewise
continuously differentiable combination of exponential 
functions, which represent an asymptotic solution of the
mathematical problem for small widths of the
region in which the transition from non-force-free to
force-free takes place.
Before applying the 
asymptotic
solution to observational data we present the results of
a
controlled test 
model
showing that the asymptotic solution substantially increase the
numerical efficiency without compromising the accuracy
of the solutions. It is
therefore a good basis for generating an efficient
numerical extrapolation tool based on the family of MHS solutions
by \citet{Neukirch2019}.

The structure of the paper is as follows. In Section 
\ref{sec:background}
we summarise the basic theory of the particular class of 3D MHS solutions by \citet{Neukirch2019}. In Section 
\ref{sec:methodology}
we present 
the new asymptotic solution and
how we test it,
the datasets used as examples,
and 
the analysis tools 
used
to 
quantitatively assess the quality of the approximation
of the exact solutions by the asymptotic solution.
Section 
\ref{sec:results}
describes
the results of our investigations, 
followed by the a final discussion in Section 
\ref{sec:summary}.

\section{Background Theory}
\label{sec:background}

As stated before, we use Cartesian coordinates with 
$z$ being the height above the solar surface. 
We follow 
the approach proposed by \citet{NeukirchRastst1999}
and use a
poloidal-toroidal decomposition \citep[e.g.][]{ChandKendall1957, Nakagawa1972}
for representing the
magnetic field,
\begin{equation} \label{eq:magfield}
\textbf{B} = \nabla \times \left[ \nabla \times \left( \Phi \hat{\textbf{z}} \right) \right] + \nabla \times \left( \alpha \Phi \hat{\textbf{z}} \right).
\end{equation}
Here, $\alpha$ is assumed to be constant, which results in 
the 
toroidal function $\Theta$ to take the form 
$\Theta = \alpha \Phi$ \citep[e.g.][]{Nakagawa1972}.

Following \cite{Low1991} we assume a current density of the form
\begin{equation} \label{eq:currentdensity}
\mu_0 \textbf{j} = \alpha \textbf{B} + \nabla \times \left(f(z)B_z \hat{\textbf{z}}\right)
= \alpha \textbf{B} + f(z) \nabla B_z \times \hat{\textbf{z}}, 
\end{equation} 
such that the non-force-free contribution to the current density in Equation \ref{eq:currentdensity} evolves according to $f(z)B_z$ with height $z$. Therefore, if $f(z)$ tends to zero for $z \to \infty$ the perpendicular part of the current density 
also tends to zero, which
leads to the magnetic field approaching a (linear)
force-free state.

The 
poloidal 
function $\Phi$ 
defining the
magnetic field  in Equation \ref{eq:magfield} is 
determined by solving Amp\`{e}re's law in the 
form \citep{NeukirchRastst1999}
\begin{equation} \label{eq:PhiPDE}
\Delta \Phi - f(z) \Delta_{xy} \Phi + \alpha^2 \Phi = 0 , 
\end{equation}
where $\Delta_{xy} \Phi = \frac{\partial^2 \Phi}{\partial x^2} + \frac{\partial^2 \Phi}{\partial y^2}$. 
Once $\Phi$ is known, the 
magnetic field can be found via Equation \ref{eq:magfield}). 

%
%
One can integrate the force balance equation 
\citep[see e.g.][]{Low1991,NeukirchRastst1999}
and obtain expressions for
the 
plasma pressure and mass density 
in the form
\begin{eqnarray}
p(x,y,z) &=& p_b(z) + \Delta p(x,y,z) = p_b(z)-f(z)\frac{B_z^2}{2 \mu_0}, \label{eq:p3Dfull}\\
\rho(x,y,z) &=& 
-\frac{1}{g}\frac{d p_b}{dz} +
\Delta \rho(x,y,z)= \frac{1}{g} \left( - \frac{d p_b}{dz} + \frac{df}{dz} \frac{B_z^2}{2 \mu_0} + \frac{f}{\mu_0} \textbf{B} \cdot \nabla B_z \right),
\label{eq:rho3Dfull}
\end{eqnarray}
where $p_b$ is 
a free function of height $z$, which arises during the 
integration of the force balance equation (an 
"integration constant").
This function
can be interpreted 
as the
plasma pressure of a stratified background 
atmosphere and while there are no mathematical restrictions on $p_b(z)$, from a physics point of view it
has to satisfy two conditions: (i) it has to
be strictly positive, i.e. $p_b(z) > 0$ for
all $z \ge 0$ and (ii) it has to be a strictly monotonically decreasing function of $z$, i.e. $\frac{d p_b}{dz} < 0$
for all $z$. The reason for these two conditions is that
this is the only way to get a positive total plasma pressure and density.
The 
plasma
temperature can then be obtained from $p$ and $\rho$ using the ideal gas law $T = \bar\mu p / (k_B \rho)$, where $\bar\mu$ is the mean atomic weight of the plasma and $k_B$ the Boltzmann constant. 

A convenient way of 
solving
the 
partial
differential equation \ref{eq:PhiPDE} for $\Phi$ 
exploits the fact that its coefficients only depend on
$z$. Therefore
one can use
Fourier 
transforms in the $x$- and $y$-coordinates, i.e.
write
\begin{equation} \label{eq:PhiIntegral}
\Phi = \iint_{- \infty}^\infty \bar{\Phi}(z; k_x, k_y) \exp \left[ i \left(k_x x + k_y y \right) \right] dk_x dk_y .
\end{equation}
The
scalar function $\bar{\Phi}$ is the solution of
the linear ordinary second-order differential equation
\begin{equation} \label{eq:PhiBarODE}
\frac{d^2 \bar{\Phi}}{dz^2} + \left[ \alpha^2 - k^2 + k^2 f(z) \right] \bar{\Phi} = 0,
\end{equation} 
where $k^2 = k_x^2 + k_y^2$. If periodic boundary conditions in $x$ and $y$ are imposed $k_x$ and $k_y$ take on discrete values and the integrals in Equation \ref{eq:PhiIntegral} 
become
sums. 

\begin{figure}
\centering
\includegraphics[width=.6\textwidth,clip=]{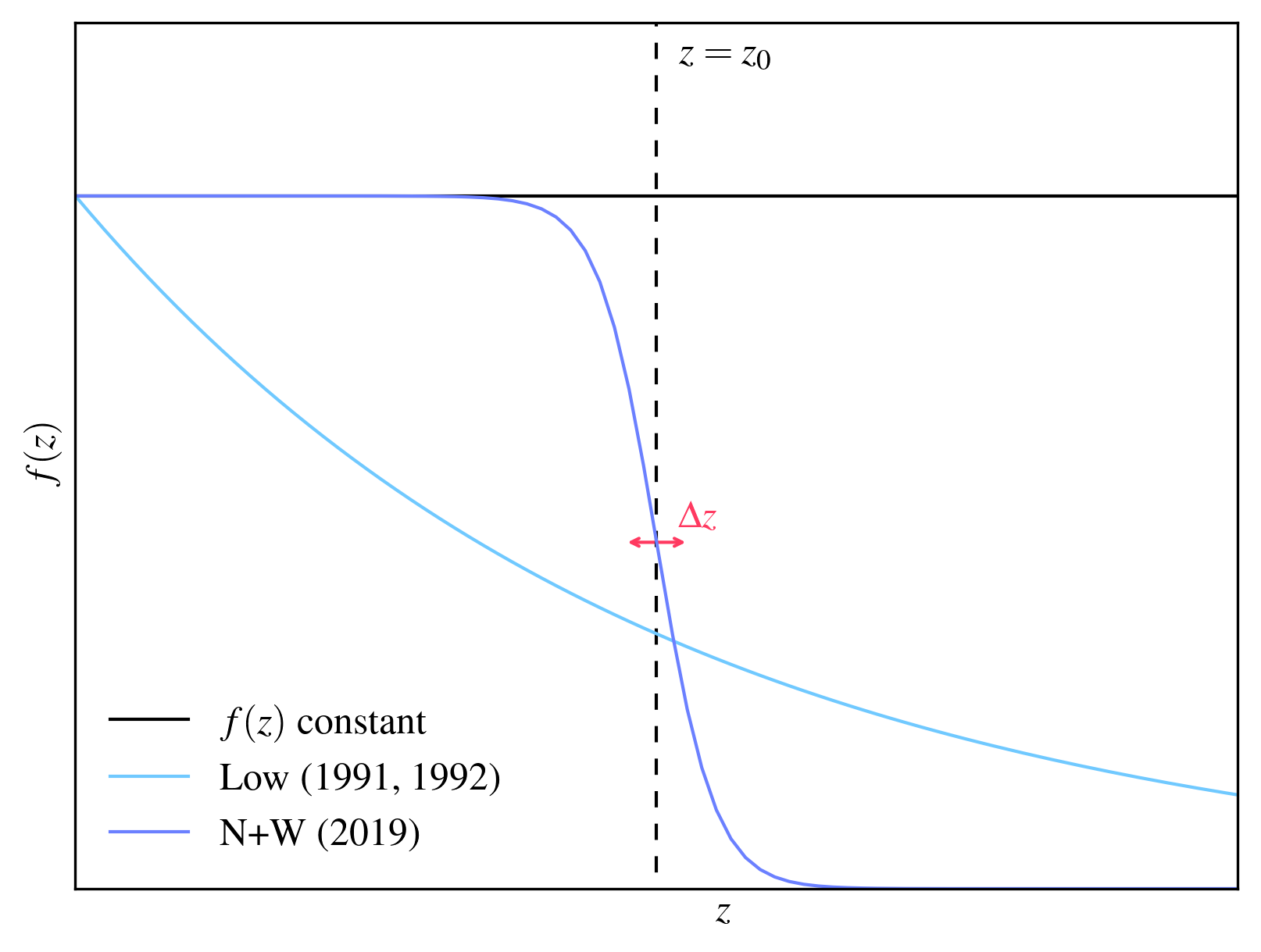}
\caption{Different choices for $f(z)$. The black line corresponds to $f(z)=$ constant, the light blue curve to the exponential function used
by \citet{Low1991}, and the dark blue curve to the hyperbolic
tangent profile from \citet{Neukirch2019}. This figure is representative of the case where $b=1.0$ in Equation \ref{eq:fNW}.}
\label{fig:f}
\end{figure}

For $f(z)=f_0 =\mbox{constant}$, the solutions
to Equation \ref{eq:PhiBarODE} are exponential functions 
\citep[e.g.][]{Neukirch1997b,Petrie2000,Petrie2000b}.
\cite{Low1991} used an exponentially decaying 
$f(z)$, i.e.
\begin{equation}
f_{L}(z) = a_{L} \exp \left(- \kappa z \right)
\label{eq:f-low}
\end{equation}
with magnitude parameter $a_L$ 
and inverse length scale
$\kappa$.
For this $f(z)$, the
solutions of Equation \ref{eq:PhiBarODE} for $\bar\Phi$ are Bessel functions. 
More recently,
\cite{Neukirch2019} 
suggested a more complicated $f(z)$ of the form
\begin{equation} \label{eq:fNW}
f_{N+W}(z) = a \left[ 1 - b \tanh \left( \frac{z - z_0}{\Delta z} \right) \right].
\end{equation}
The corresponding
solutions 
for 
$\bar\Phi$ 
can be expressed in terms of
hypergeometric functions. 

These three different choices for the function $f(z)$ are shown in Fig. \ref{fig:f} for a specific combination of parameters in Equation \ref{eq:fNW} ($b=1.0$, $a_L = a (1- \tanh(-z_0 / \Delta z)$, $\kappa = 1/z_0$, and the constant $f(z)$ such that it matches $f_L$ and $f_{N+W}$ on the photosphere). As one can see in Fig. \ref{fig:f}, if the
argument of the hyperbolic tangent function in 
$f_{N+W}(z)$ is large and negative the function
is close to the asymptotic value $a(1+b)$, whereas in tends to the value $a(1-b)$ if the argument of
the hyperbolic tangent is large and positive.
Therefore, the parameter
$a$ controls the overall magnitude of 
$f_{N+W}$, whereas $b$ controls the difference between
the two asymptotic values of $f_{N+W}(z)$.
In the case $b=1$, the parameter $b$ can be considered 
the "switch off" parameter, because the asymptotic value
of $f_{N+W}(z)$ for $(z-z_0)/ \Delta z \gg 0$ is zero 
in this case, implying that the magnetic field tends towards
a linear force free field in this limit. The 
example for $f_{N+W}(z)$ shown in Fig. \ref{fig:f} is for $b=1$.
As Fig. \ref{fig:f} also shows,
$z_0$ is the central height around 
which the transition from non-force-free to 
force-free magnetic field occurs and 
$\Delta z$ controls the width over 
which that transition takes place. 
If $b \neq 1$ a certain degree of non-force-freeness can
be maintained in the upper part of the model above $z_0$
\citep{Neukirch2019}. 
We remark that a similar effect could be introduced 
into $f_L(z)$ by adding
a constant parameter $b_{L}$ to the exponential function.
The $f_{N+W}(z)$
profile 
has
more
parameters than the $f_L(z)$ function.
The aim is
to have
more control over the transition from the non-force-free to
force-free domain and,
as a consequence,
more control 
over the resulting 
pressure and density profiles. 
%

The solution of Equation \ref{eq:PhiBarODE} with Equation \ref{eq:fNW} by \cite{Neukirch2019} is given by
\begin{equation} \label{eq:PhiBarNW}
\bar{\Phi}_{N+W} = \bar{A} \eta^{\bar{\delta}} \left(1-\eta\right)^{\bar{\gamma}} 
{}_2F_1 \left( \bar{\gamma} + \bar{\delta} + 1, \bar{\gamma} + \bar{\delta}, 2 \bar{\delta} 
+ 1; \eta \right), 
\end{equation}
where ${}_2F_1(a, b, c; z)$ is the hypergeometric function\footnote{See NIST Digital Library of Mathematical Functions, \url{https://dlmf.nist.gov/15}.}. 
Here, the argument $\eta$ is defined as
\begin{equation}
    \eta = \frac{1}{2}\left[1-\tanh\left(
    \frac{z-z_0}{\Delta z}\right) \right],
    \label{eq:etadef}
\end{equation}
and 
the parameters are given by the expressions
$\bar{\gamma} = \sqrt{C_2}$ and $\bar{\delta} = \sqrt{C_1}$, 
with
\begin{align*}
C_1 &= \frac{1}{4} \left[ \bar{k}^2 \left( 1 - a + ab \right) - \bar{\alpha}^2 \right], \\
C_2 &= \frac{1}{4} \left[ \bar{k}^2 \left( 1 - a - ab \right) - \bar{\alpha}^2 \right].
\end{align*}
Here
$\bar{k} = k \Delta z$ and $\bar{\alpha} = \alpha \Delta z$ 
\citep{Neukirch2019}, and $\tilde{A}$ is determined 
by the boundary 
conditions. We remark that in Equation \ref{eq:PhiBarNW} we have only 
included the part of the general solution which tends to zero
as $z \to \infty$ 
\citep[for the full solution, see][]{Neukirch2019}. 

%
%
\section{Methodology}
\label{sec:methodology}


\subsection{Approximate solutions in the asymptotic regime}

%
%

The solution presented in Equation \ref{eq:PhiBarNW} for the case when $f_{N+W}$ is used in the ODE \ref{eq:PhiBarODE} is 
general, but it still presents some challenges due to the hypergeometric functions involved.
Although hypergeometric functions are a standard class of special functions, it is not necessarily 
intuitively clear how their general behaviour changes depending on the parameter values used. Furthermore,
the hypergeometric differential equation has singular points and it turns out that the solution \ref{eq:PhiBarNW}
can approach one of these singular points. This can be problematic for the numerical evaluation 
of the hypergeometric functions with respect to both accuracy and speed
even when using well-tested numerical implementations as, for example, the Python SciPy library.
However, it turns out that one can use the properties of $f_{N+W}(z)$ and the resulting asymptotic
behaviour of the
ODE \ref{eq:PhiBarODE} in a certain parameter regime to obtain an approximate solution that 
can be expressed purely in terms of exponential 
functions. It is one of the main aims of this paper to present the derivation of this approximate solution in
the asymptotic regime and to assess its accuracy in representing the exact solution. This, together with
an assessment of the gain in computational speed, will allow us to evaluate whether and when this approximate
solution can be safely used to replace the exact solution.


%
%
We define $\tilde{z} = z/L$, $\tilde{\alpha} = \alpha L$,
$\tilde{k} = k L$, and $\Delta \tilde{z}= \Delta z/L$, with $L$ a 
general normalising length scale. The relation to the previous normalisation with 
$\Delta z$ as normalising length scale is simply given by 
$ \bar{\alpha} = \tilde{\alpha} \Delta \tilde{z}$, $ \bar{k} = \tilde{k} \Delta \tilde{z}$, and so on.

%
%
As already indicated in Section \ref{sec:background}, the function $f_{N+W}(z)$ asymptotically tends 
to constant values for large negative or positive argument of the hyperbolic tangent function (i.e. for $|z-z_0| \gg \Delta z$ in Equation \ref{eq:fNW}. 
%
If $\Delta \tilde{z}$ is small the transition between the asymptotically constant values of
$f_{N+W}$ takes place over a very small domain in $\tilde{z}$, centred at $\tilde{z}_0$.
%
%
In the domains where $f_{N+W}$ tends to its asymptotic values equal to $f \simeq a(1 \pm b)$, 
Equation \ref{eq:PhiBarODE} is reduced to
\begin{equation} 
\label{eq:PhiBarODEAsymp}
\frac{d^2 \bar{\Phi}}{d\tilde{z}^2} + \left[ \tilde{\alpha}^2 - \tilde{k}^2 \left( 1 - a \pm ab \right) \right] \bar{\Phi} = 0,
\end{equation}
with the $+$ sign applying for $\eta = (\tilde{z} - \tilde{z}_0)/\Delta\tilde{z} > 0$ and the $-$ sign
for $\eta < 0$.
We define
\begin{equation*}
C_{\pm} =\frac{1}{4}\left[  \tilde{k}^2 \left( 1 - a \pm ab \right) -\tilde{\alpha}^2  \right]
\label{eq:Cpm-def}
\end{equation*}
and assume that $C_\pm> 0$ (we note that $C_1 = C_+ \Delta \tilde{z}^2 $ and $C_2 = C_- \Delta \tilde{z}^2 $). 
The above asymptotic limit is equivalent to using the alternative function
\begin{equation*}
    f(z) = \begin{cases}
        a (1-b) &\text{for } z < z_0 \\
        a (1+b) &\text{for } z > z_0
    \end{cases}
\end{equation*}
instead of Equation \ref{eq:fNW}. This dicontinuous $f(z)$ introduces a jump in Equation \ref{eq:PhiBarODE}, yielding Equation \ref{eq:PhiBarODEAsymp}.

%
%
An approximation to the exact solution $\bar{\Phi}$ of Equation \ref{eq:PhiBarODE} can be found by solving Equation \ref{eq:PhiBarODEAsymp} in the two domains
$0 \le \tilde{z} \le \tilde{z}_0$ and $\tilde{z}_0 < \tilde{z} < \infty$, under the following conditions
%
%
\begin{enumerate}
\item $\bar\Phi$ 
is once continuously differentiable at $\tilde{z} = \tilde{z}_0$,
%
%
\item $\bar\Phi \to 0$ as $\tilde{z} \to \infty$, and
\item $\bar\Phi = 1$ at $z=0$.
\end{enumerate}
%
%
Defining $\delta = \sqrt{C_+}$ and $\gamma = \sqrt{C_-}$, the solution is given by
\begin{equation} \label{eq:PhiBarAsymp}
\bar{\Phi} = \frac{1}{D} \begin{cases} 
\frac{\delta}{\gamma} \sinh \left[ 2 \gamma (\tilde{z}_0-\tilde{z}) \right] + 
\cosh \left[ 2 \gamma (\tilde{z}_0-\tilde{z}) \right] & \mbox{for } 0 \le \tilde{z} \le \tilde{z}_0   \\
\exp \left[ - 2 \delta (\tilde{z}-\tilde{z}_0) \right] & \mbox{for } \tilde{z}_0 < \tilde{z} < \infty
\end{cases},
\end{equation}
with 
\begin{equation*}
D = \frac{\delta}{\gamma} \sinh \left( 2 \gamma z_0 \right) + \cosh \left( 2 \gamma z_0 \right).
\end{equation*}
In Equation \ref{eq:PhiBarAsymp}, we
have for ease of notation suppressed the dependence of $\bar{\Phi}$
on the value of $k^2$, linking it to the specific Fourier modes.
We will use this simpler notation for the remainder of this paper.

The first derivative of $\bar{\Phi}$ is
\begin{equation} \label{eq:DPhiBarAsymp}
\bar{\Phi}' = - \frac{1}{D} \begin{cases} 
 2 \delta  \cosh \left[ 2 \gamma (\tilde{z}_0-\tilde{z}) \right] + 
 2 \gamma \sinh \left[2 \gamma (\tilde{z}_0-\tilde{z})\right] & \mbox{for } \le \tilde{z} \le \tilde{z}_0  \\
  2 \delta \exp \left( - 2 \delta (z-z_0) \right) & \mbox{for } \tilde{z}_0 < \tilde{z} < \infty
\end{cases}.
\end{equation}

\begin{figure}
\centering
\includegraphics[width=0.49\textwidth,clip=]{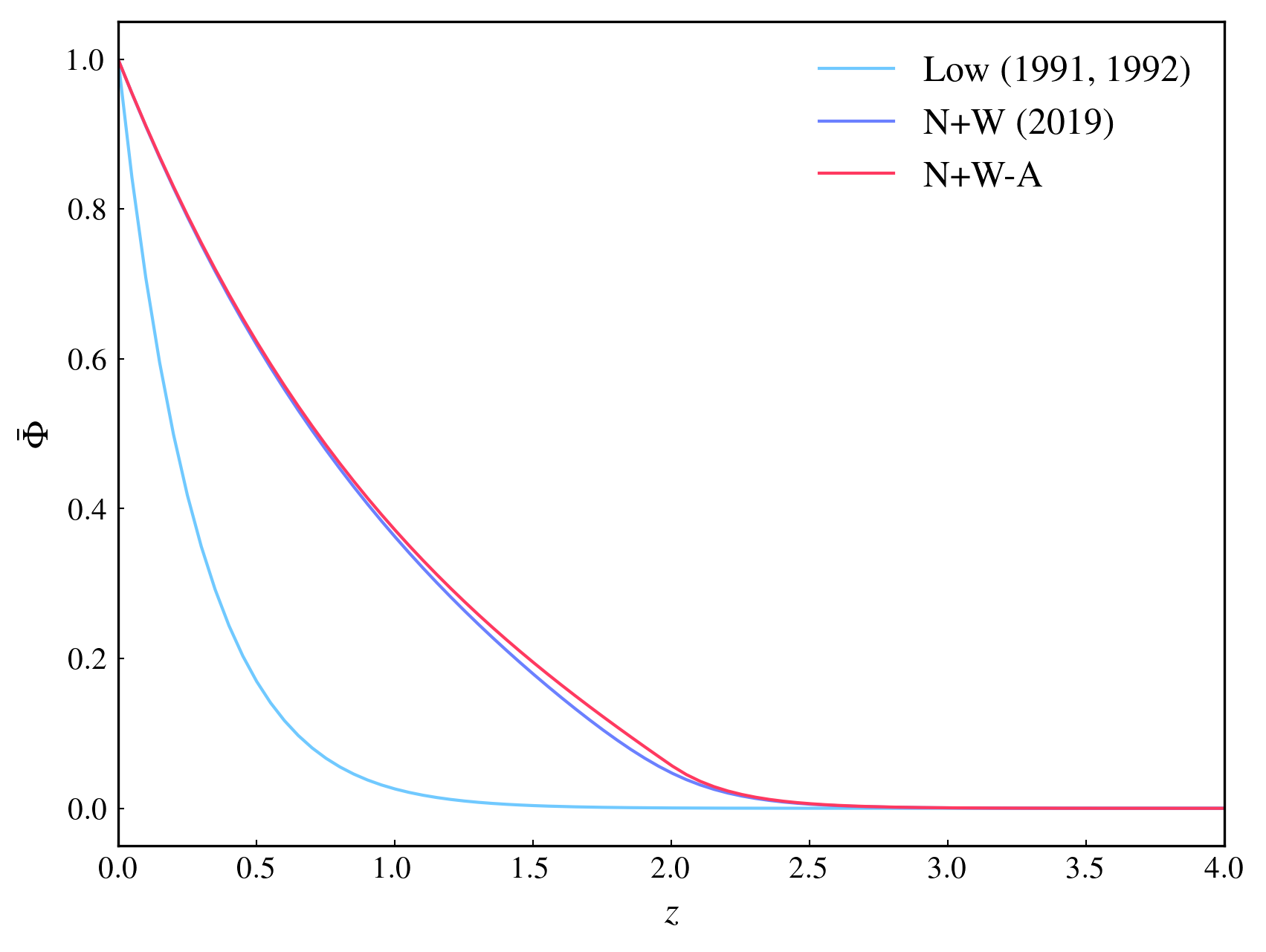}
\includegraphics[width=0.49\textwidth,clip=]{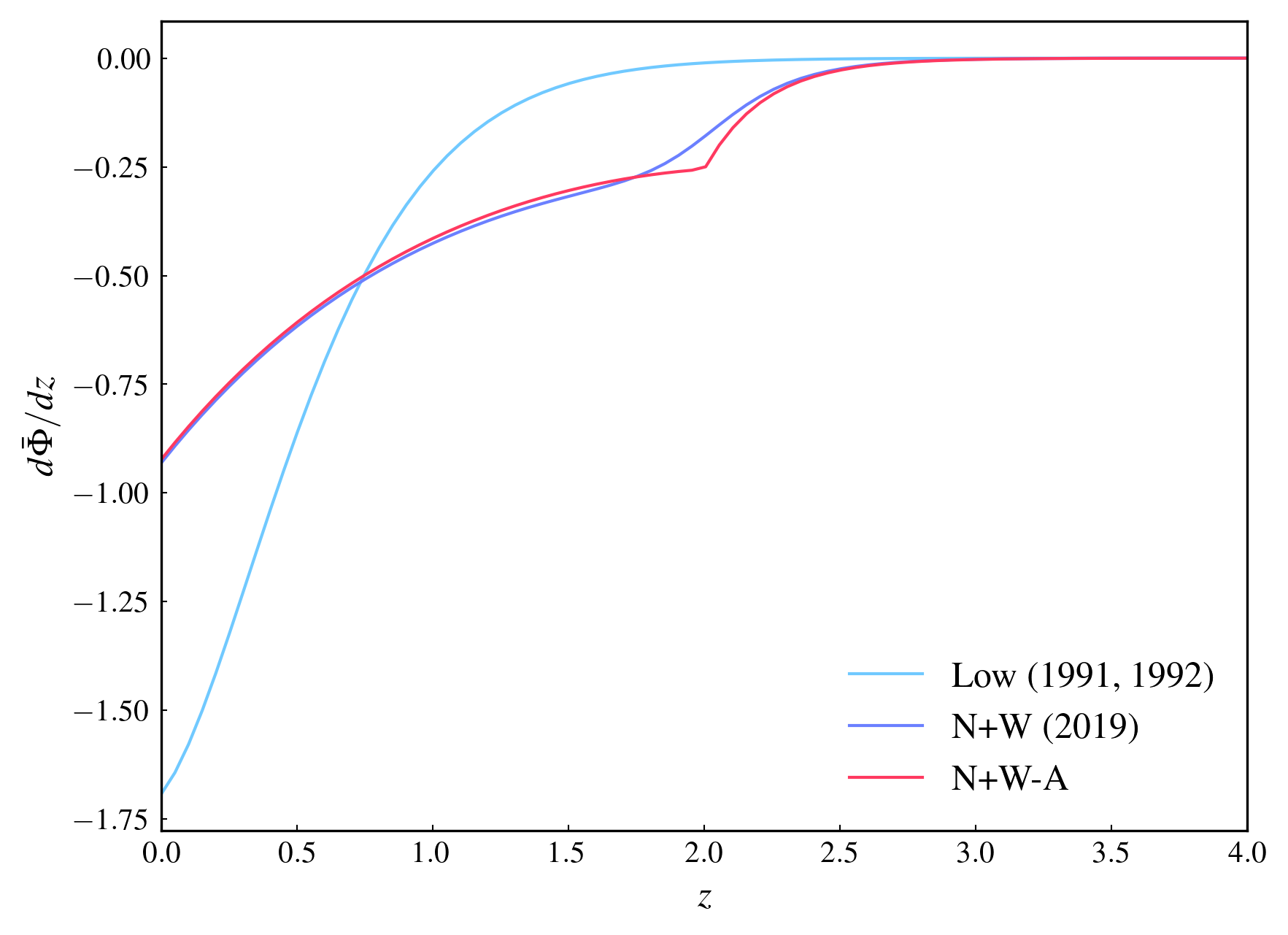}
\caption{
%
%
$\bar\Phi$ (left panel) and $d \bar\Phi / d\tilde{z}$ (right panel) for a single Fourier mode
($\tilde{k}^2 = 2\pi^2\approx 19.7392$). Shown are the exact 
solution (labelled N+W (2019), dark blue), the approximate asymptotic solution (labelled N+W-A, red), and
for comparison a solution based on $f_L(z)$ (labelled Low (1991, 1992), light blue).
The parameter values used for the N+W and N+W-A curves are $\tilde{z}_0 = 2.0$,
$\Delta \tilde{z} = 0.2$, $a=0.48$, $b=1.0$ and $\tilde{\alpha} = 0.03$. The parameter values used for the 
Low (1991, 1992) curve are $\kappa = 1/\tilde{z}_0$ and $a_L = a (1+\tanh(\tilde{z}_0 / \Delta \tilde{z}))$.
All functions are shown over the domain $0 \le \tilde{z} \le 2\tilde{z}_0$ to include not only
the region around $\tilde{z}_0$ where $f_{N+W}$ changes between its asymptotic values, but also parts of
the regions outside this transitional domain. It is obvious from this plot that at least in this particular
case the asymptotic solution and its first derivative are very good approximations of their exact counterparts.
Unsurprisingly, the Low (1991, 1992) curve deviates significantly from both.
%
}
\label{fig:phi}
\end{figure}

The behaviour of this approximate solution based on the asymptotic form of the ODE  \ref{eq:PhiBarODEAsymp}
%
%
is consistent with the asymptotic behaviour of the exact solution Equation \ref{eq:PhiBarNW}, as 
found analytically 
by
\cite{Neukirch2019}. 
An illustrative example for a typical combination of parameter values and a single value of $\tilde{k}^2$,
i.e.\ a single Fourier mode, is
shown in
Figure \ref{fig:phi}. The figure displays plots of
three different functions. In the left panel, we show
the
asymptotic solution derived above (labelled N+W-A), the exact solution 
(labelled N+W), and for comparison a solution based on
$f_L(z)$ defined in Equation \ref{eq:f-low}, using parameter values that are derived from the parameter values
used for $f_{N+W}$ (for details we refer to the caption of Figure \ref{fig:phi}).
The panel on the right shows plots of the first derivatives of these three functions.
%
%
The asymptotic solution approximates the exact solution generally very well. As is to be expected, the largest
deviations of the asymptotic solution from the exact solution can be seen in the region 
around $\tilde{z} \approx \tilde{z}_0$, and it is more pronounced in the derivative (right panel) than in the
function itself.
Not surprisingly, the \cite{Low1991, Low1992} solution differs significantly from the other two solutions,
despite matching the parameter values for $f_L(z)$ and $f_{N+W}$ as much as possible.

%
%
Although this illustrative example is promising,
an assessment of the quality of the approximation should not be based on a single Fourier mode and a specific parameter set. 
One of the major purposes of this paper is
to present a 
detailed comparison between magnetic
field extrapolations based on the exact solution and the asymptotic solution, and how the 
differences between the two depend on the parameters of $f_{N+W}$, in particular $a$ and $\Delta \tilde{z}$.
This will allow a much more solid assessment of the quality of the approximation provided 
by the asymptotic solution.
In this paper we will only investigate the case $b=1$, which means that the magnetic field very quickly
approaches a linear force-free state above $\tilde{z}_0$.

The following considerations allow us to restrict the parameter
range to sensible values. Firstly, we shall use the
method by \citet{Seehafer1978} to deal with
unbalanced magnetograms. This implies that periodic 
boundary conditions are imposed in the
$x$- and $y$-directions on the extended domain used by 
the Seehafer method. Hence, the solution of Equation 
\ref{eq:PhiPDE} is given by a
Fourier series instead of a Fourier transformation as in
Equation \ref{eq:PhiIntegral}. We define the wave
number associated with the fundamental mode
of this Fourier series as $\tilde{k}_{min}$. Following
\citet{Neukirch2019}, we assume that $C_- > 0$ (this also
implies $C_2 > 0$) so that the parameter $\gamma$
(and $\bar{\gamma}$) only takes on real values. We note
that due to $C_+ \ge C_-$ ($C_1 \ge C_2$) 
this also ensures that $\delta$ only takes on real values.
As shown by \citet{Neukirch2019}, this condition means
that
the
maximum possible value for $a$ is given by\footnote{
In turns out that this range can in principle be extended 
to allow values above $a_{max}$, but we do not investigate this case here due to the unclear mathematical
complications associated with these cases.} 
\begin{equation}
     a_{max} = \frac{1}{b+1}\left[1 - \left(\frac{\tilde{\alpha}}{\tilde{k}_{min}}\right)^2\right].
    \label{eq:amax}
\end{equation}
We also impose $\tilde{\alpha}^2 < \tilde{k}_{min}^2$ to
keep $C_+ > 0$ (for the case $b=1$).

%
%
%
The second important model parameter whose influence on the accuracy of the asymptotic solution 
has to be investigated is $\Delta \tilde{z}$. As already stated before, the idea behind the asymptotic solution
is that $\Delta \tilde{z}$ is small, so that for most of the domain in $\tilde{z}$ the ODE \ref{eq:PhiBarODE}
is well approximated by the asymptotic ODE \ref{eq:DPhiBarAsymp}. In other words, the value of $\Delta \tilde{z}$
has to be small enough so that hyperbolic tangent function can be considered to be replaced by
a step function in the 
ODE \ref{eq:PhiBarODE}. We emphasise that this is only done to calculate $\bar{\Phi}_{N+W-A}$ and the
resulting magnetic field, but that the exact form of $f_{N+W}(z)$ is used in the calculation
of the pressure and the density. We discuss the reason for this below.
%
%
The obvious question is which other length scale we compare $\Delta \tilde{z}$ with to decide whether it
is small. Given that
the argument of the hyperbolic tangent function in $f_{N+w}$ is $\eta = (\tilde{z} - \tilde{z}_0)/\Delta \tilde{z}$,
a natural choice for such a comparison is $\tilde{z_0}$. In this paper, we will therefore test the 
quality of the asymptotic solution only up to an upper bound of $\Delta \tilde{z} = \tilde{z}_0$.
%
Because $\Delta \tilde{z}$ has to be small for the asymptotic solution to be accurate, it seems tempting
to consider the limit $\Delta \tilde{z} \to 0$. In this case, the hyperbolic tangent would become a step function,
and the asymptotic solution would actually become the exact solution. However, this is not a
physically valid limit due to the expression for the density, given in Equation \ref{eq:rho3Dfull}.
This expression depends on the first derivative of $f$ with respect to $z$. In the limit $\Delta \tilde{z} \to 0$ this 
derivative turns into a Dirac $\delta$-function at $\tilde{z} = \tilde{z}_0$, 
leading to the density going to (negative) infinity at that point, which is clearly unacceptable.
One therefore also needs to impose a finite lower bound on $\Delta\tilde{z}$ which is on 
the one hand small enough to make the asymptotic solution sufficiently accurate, but
which on the other hand keeps the transition between the two asymptotic regions smooth enough. 
There is also another reason why the $\Delta \tilde{z}$ value cannot be chosen too small. It turns
out that very small values for $\Delta \tilde{z}$ can cause numerical problems with the calculation of
the exact solution. Hence we choose  
$\Delta\tilde{z}= 0.1 \tilde{z}_0$ as the typical value. We remark that in this case if one takes $L = 10^6 $m and 
$\Delta{z}_0 = 2.0$
(i.e. $z_0 = 2,000$km) one obtains a reasonable value for 
the width of the transitional region, namely $\Delta z = 200$ km. 




%
%
\subsection{Boundary conditions for test cases}

\label{sec:Data}

To assess the quality of the approximation of the exact solution 
by the asymptotic solution,
we 
consider 
two different photospheric boundary conditions for the magnetic field component $B_z$.
As already mentioned before we use the method by \citet{Seehafer1978} to deal 
with boundary data for which the magnetic flux through the lower boundary ($\tilde{z}=0$) is
unbalanced. This implies periodic boundary conditions 
(on the extended domain required by the method) in the $x$- and $y$-directions.

The first boundary condition is given analytically and represents a generalisation to a multipole of the
analytical periodic bipole boundary condition used by \cite{Neukirch2019}. 
The second boundary condition uses magnetogram data taken 
by the Solar Dynamics Observatory \citep[SDO,][]{Scherrer2012}.
This case
tests whether
the asymptotic method can be successfully applied 
to observational data, 
providing
a more realistic 
assessment of the 
quality of the asymptotic solution method. 
Both cases also 
allow
us to test the 
gain in numerical efficiency by using the much simpler asymptotic solution in comparison with
the exact solution. 


\subsubsection{Analytical multipole boundary conditions} \label{sec:AnMu}

\begin{figure} 
\centering
    \includegraphics[width=0.5\textwidth]{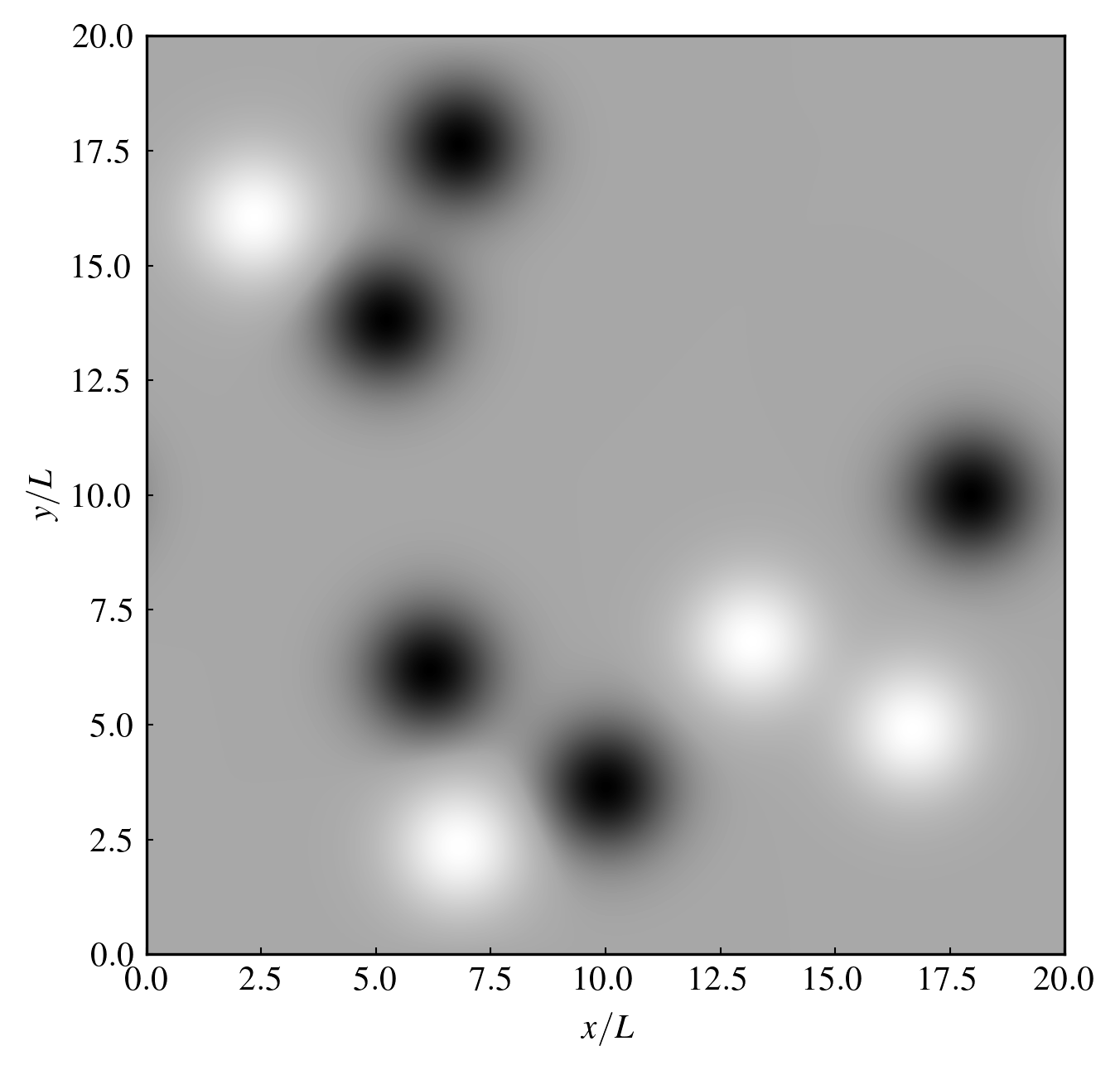}
    \caption{Artificial line-of-sight magnetogram created from Equation \ref{eq:VonMises} using the parameters given in the text as well as in Table \ref{tab:art-params}.}
    \label{fig:AnaMulContourf}
\end{figure}

\begin{table}[]
    \caption{Parameter values for the analytical
    multipole boundary condition case, 
    Equation \ref{eq:VonMises}.}
    \label{tab:art-params}
    \begin{tabular}{ccccc}
        \hline
      $i$  & $\mu_{x,i}$ & $\mu_{y,i}$ & $\lambda_{x,i}$ & $\lambda_{y,i}$ \\
         \hline 
         \rowcolor{lavender}
      0  &  1.0 & -1.0 &  10.0 & 10.0\\   
      1  & -1.2 & -1.2 &  10.0 & 10.0\\
      \rowcolor{lavender}
      2  & -2.4 &  1.9 &  10.0 & 10.0\\
      3  &  2.1 & -1.6 &  10.0 & 10.0\\
      \rowcolor{lavender}
      4  & -1.5 &  1.2 &  10.0 & 10.0\\
      5  &  2.5 &  0.0 &  10.0 & 10.0\\
      \rowcolor{lavender}
      6  &  0.0 & -2.0 &  10.0 & 10.0\\
      7  & -1.0 & -2.4 &  10.0 & 10.0\\
      \rowcolor{lavender}
      8  & -1.0 &  2.4 &  10.0 & 10.0\\
        \hline
    \end{tabular} 
\end{table}
The
artificial boundary condition 
we use to test
our method is
constructed similarly to the bipole boundary condition in \cite{Neukirch2019}. 
The 
vertical magnetic field component 
at $z=0$ is given by
\begin{equation} \label{eq:VonMises}
B_z(x,y,0) = B_0 \sum_{i=0}^8 (-1)^{i+1} 
%
\frac{\exp(\lambda_{x,i} \cos(\hat{x}-\mu_{x,i}))}{2 \pi \text{I}_0(\lambda_{x,i})} 
\frac{\exp(\lambda_{y,i} \cos(\hat{y}-\mu_{y,i}))}{2 \pi \text{I}_0(\lambda_{y,i})}
\end{equation}
%
%
where $\text{I}_0$ is a modified Bessel functions of the first kind.
To achieve a compact form of the argument of the cosine function in the exponentials in Equation \ref{eq:VonMises}
we have assumed that the region under consideration on the bottom boundary is $0\le x \le x_0, 0\le y \le y_0$.
We then define
$\hat{x} = \pi \left( 2x/x_0- 1\right)$ and 
$\hat{y} = \pi \left(2y/y_0-1 \right)$, so that $-\pi\le \hat{x}, \hat{y} \le \pi $.

The terms in 
Equation\
\ref{eq:VonMises} 
with odd $i$ 
%
%
%
correspond to
regions of positive magnetic polarity
and the ones with even $i$ 
correspond to
regions of negative magnetic polarity.
Because the sum has nine terms,
the resulting magnetogram is not flux-balanced. We made this choice to have a controlled test of our 
implementation of the \citet[][]{Seehafer1978} method.
The dimensionless parameters $\lambda_{x,i}$ and $\lambda_{y,i}$ can be regarded
as normalised inverse length scales, which determine the width of each magnetic flux
region
in the $x$- and $y$-direction. 
We have chosen a uniform value of $\lambda_{x,i} = \lambda_{y,i} = \lambda= 10.0 $. 
The parameters 
$\mu_{x_i}$ 
and $\mu_{y_i}$
determine the positions of the centres of the magnetic flux sources within the 
computational domain.
The specific values of all model parameters chosen for this case are listed in Table \ref{tab:art-params}. 
The resulting photospheric $B_z$ is shown in Figure \ref{fig:AnaMulContourf}.
In normalised coordinates, the computational 
domain
has an
extent of 
%
%
$20.0$ 
in each direction.
We choose $z_0 = 2.0$, which for a normalising length scale $L=1$ Mm corresponds to a height of $2.0$ Mm,
which approximately corresponds to the height of the transition region in the solar atmosphere.

%
%
\subsubsection{Observational data boundary conditions} \label{sec:Obs}

\begin{figure} 
\centering
    \includegraphics[width=0.8\textwidth]{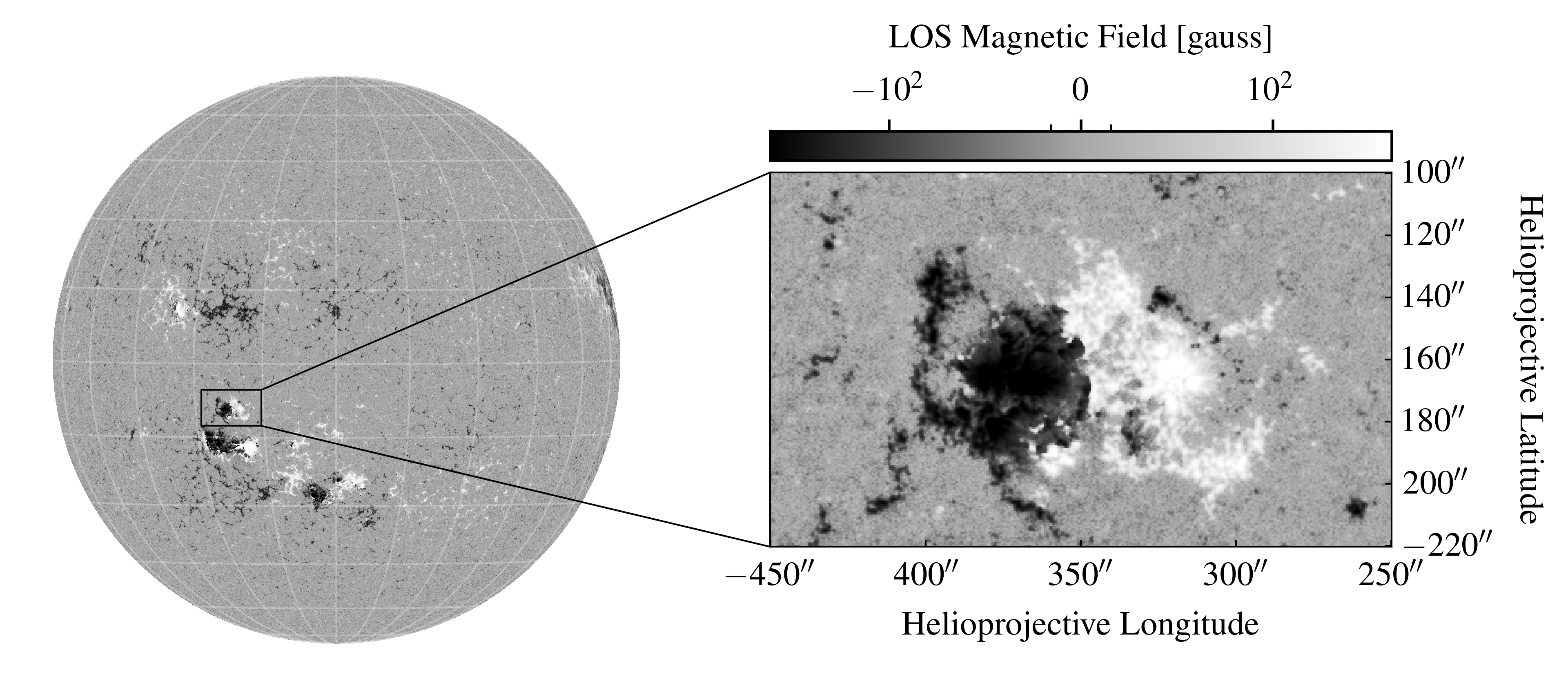}
    \caption{Full disk line-of-sight magnetogram observed by HMI onboard SDO 
    on 2012 June 13 at 07:31:30 UT with zoomed in cutout of the local active region used for the application of the 
    extrapolation method based on the exact solution family
    by \cite{Neukirch2019} and its asymptotic approximation.}
    \label{fig:Obs}
\end{figure}

In this paper we aim to apply the \citet[][]{Neukirch2019} MHS solutions family and the asymptotic
approximation derived above to observational data for the first time.
For this, 
we 
use a photospheric line-of-sight magnetogram taken by the 
Helioseismic Magnetic Imager \citep[HMI, ][]{Schou2012} onboard SDO. 
We choose an observation from 2012 June 13 at 07:31:30 UT as an example case for this study. 
The full disk line-of sight HMI magnetogram is shown in Figure \ref{fig:Obs} together with the
zoomed-in cutout ($397 \times 239$ pixel) we use as boundary condition for magnetic field extrapolation.
As one can see the cutout shows a largely bipolar magnetic polarity structure with some smaller substructure
visible as well. This magnetogram 
was chosen because it
will provide a very good test for our MHS extrapolation method in general, and
for a comparison of the exact and asymptotic solution cases.


As the active region under investigation here is close to the centre of the disk no distinction has been made between $B_{LOS}$ and the radial field component $B_r$ in the heliographic projection, which should theoretically be used. Therefore, for the purpose of this paper, which is to compare the exact to the analytical solution, the differences between $B_{LOS}$ and $B_r$ are ignored.    

\subsection{Analysis tools}
\label{sec:analysistools}
%
%



Because the MHS solution for the magnetic field $\mathbfit{B}$ is given analytically, it can
in principle be calculated at every position $(x, y, z)$ within the computational domain.
However, calculating the magnetic field and any other solution quantities only on a fixed discrete grid is
advantageous for various reasons. It is, for example, numerically much more efficient to
generate field line plots by using magnetic field values stored on a grid and use an interpolation technique than
to reevaluate the Fourier sums by which the magnetic field is defined every time. 

As will be discussed below, we will use a number of metrics to assess the quality of the
asymptotic MHS solution in comparison with the exact MHS solution. This is also done 
more efficiently
by using values
on a grid. Finally, using a grid with a given number of grid points makes the assessment of the
computational efficiency (run time) of the numerical code much easier.

Another, related point is the number of Fourier modes used in the calculation of the magnetic field and 
any derived quantities. For the analytical multipole boundary condition, all Fourier coefficients are
in principle
given analytically. 
Nevertheless, the Fourier series has to be truncated at a finite number of modes
for computational evaluation. Usually, the truncation point has to be chosen in such a way that the Fourier series
has sufficiently converged. However, because the intention of this paper is to test the numerical
extrapolation method, we will actually generate artificial magnetograms from the analytical boundary
conditions and use these in the same way as magnetograms based on observations.
For boundary conditions based on 
magnetograms have a finite number of pixels and the resolution of the magnetogram determines the 
maximum number of Fourier modes that should be used.


As briefly mentioned above we will use standard 
metrics, sometimes also called "figures of merit", \citep[e.g.][]{Schrijver2006, Derosa2009}
to carry out a quantitative assessment regarding the quality of the approximation
of the exact magnetic field
solution by the asymptotic solution.
We also include a metric to compare the pressure
and density in both solutions.
To evaluate these metrics we assume that we have two 
magnetic
fields, 
$\mathbfit{B}_1$ and $\mathbfit{B}_2$, 
which we want compare. 
We assume that these 
magnetic
fields 
are given on a finite three-dimensional 
uniform and homogeneous 
numerical grid with consecutively numbered grid 
points labelled
by an index $i$. Here
$1 \le i \le N$, with $N$ being 
the total size of our grid, i.e. 
$N=n_x \cdot n_y \cdot n_z$, where $n_x$, $n_y$, $n_z$
are the grid resolution in the $x$-, 
$y$- and $z$-directions, respectively.
%
The metrics used are the following:
\begin{enumerate}
	\item[(1)]	The vector correlation metric
    \begin{equation*}
		C_{Vec} = \frac{\sum\limits_i^N \mathbfit{B}_{1, i}\cdot 
        \mathbfit{B}_{2, i}}
        {\left(\sum\limits_i^N |\mathbfit{B}_{1, i}|^2 
         \sum\limits_i^N |\mathbfit{B}_{2, i}|^2 \right)^{1/2}}
	\end{equation*}
	compares the local characteristics of the field vectors, where a value of $0$ indicates no correlation and $1$ is achieved by identical vectors.
	\item[(2)]	The Cauchy-Schwarz metric
	\begin{equation*}
        C_{CS} = \frac{1}{N} \sum\limits_i^N 
        \frac{\mathbfit{B}_{1, i} \cdot 
              \mathbfit{B}_{2, i}}
        {|\mathbfit{B}_{1, i}|
         |\mathbfit{B}_{2, i}|}
	\end{equation*} 
	measures the angle between the two fields based on the Cauchy-Schwarz inequality. 
    The value of $C_{CS}$ ranges from $-1$ (antiparallel fields) to $1$ (parallel fields), 
    with $0$ being assigned to perpendicular fields.
	\item[(3)]	
    Two simple measures of the difference between the two vector fields
    are
    the normalised vector error metric
	\begin{equation*}
		E_n = 
         \frac{\sum\limits_i^N |\mathbfit{B}_{1, i} - \mathbfit{B}_{2, i}| }{
                \sum\limits_i^N |\mathbfit{B}_{1, i}|
        },
	\end{equation*}
	which is the mean error normalised by the average vector norm, and
	\item[(3)]	the mean vector error metric
	\begin{equation*}
		E_m = 
        \frac{1}{N} \sum\limits_i^N \frac{|\mathbfit{B}_{1, i} - \mathbfit{B}_{2, i}|}
        {|\mathbfit{B}_{1, i}|},
	\end{equation*}
	which is the mean error 
    divided by the total number of grid points.
    The best 
    agreement between the two magnetic fields
    is achieved for $E_n = E_m = 0$.
	\item[(4)] The magnetic energy metric
	\begin{equation*}
		\varepsilon = \frac{\sum\limits_i^N \mathbfit{B}_{2, i}^2}{\sum\limits_i^N\mathbfit{B}_{1, i}^2}
	\end{equation*}
	puts the reconstructed magnetic energy in relation to the reference magnetic energy. 
    Identical magnetic energies 
    lead 
    to $\varepsilon = 1$. 
	\item[(5)] 
    A slightly different type of metric is the field line divergence metric, $l_{Div}$.   
    To calculate $l_{Div}$ we trace field lines from a random point on the bottom boundary 
    using both 
    $\mathbfit{B}_1$ and $\mathbfit{B}_2$.
    If both field lines end again on the bottom boundary, 
    a score $p_i$ 
    is
    given to this point 
    which is defined as
    the distance between the two endpoints divided by the length of 
    the field line 
    of one of the magnetic fields, say $\mathbfit{B}_1$.
    A value can then be assigned to $l_{Div}$ by working out the fraction of the area in which
    $p_i$ is below a certain threshold. We follow \cite{Zhu2022} who chose a threshold of 10 $\% $ ($0.1$).
    Obviously, the threshold value is somewhat arbitrary, for example a value of 20 $\%$ ($0.2$) was used by
    \cite{zhu2019}.
    With a threshold of 10 $\%$, a value of $l_{Div}=1$ 
    implies that
    all tested and 
    closed field lines 
   end 
    in a proximity of 10 
   $\%$
    of their length to one another. 
	\item[(6)] 
    To compare the pressure and density in the two
    solutions we use the standard linear Pearson correlation coefficients 
    for the line-of-sight 
    integrated
    (i.e. with respect to $z$)
    pressure ($r_P$) and density ($r_D$): a score between 0 and 1 indicates positive 
    correlation, a score between -1 and 0 indicates negative correlation and $r_P = r_D =0$ indicates no 
    correlation.
\end{enumerate} 
For consistency with the other metrics, 
we will use $1-E_n$ and $1-E_m$  instead of $E_n$ and $E_m$. 
The best 
agreement between the solutions
is then indicated by a value of 1
across all metrics. 

\section{Results}
\label{sec:results}

%
%

Having discussed our methodology in the previous section, we will 
now proceed to assess whether using the asymptotic MHS solution
instead of the exact solution is (a) sufficiently accurate, and
(b) leads to an adequate computational speed-up of the
method. We will start with the analytical multipole boundary
condition to test the effect of varying the parameters 
$\Delta \tilde{z}$ and $a$, while keeping the values
of the parameters $\tilde{z}_0$, $\tilde{\alpha}$ and $b$ fixed.
After that we will proceed to apply the extrapolation code
to the SDO/HMI magnetogram discussed in Section \ref{sec:Obs}.
For this case we will compare the results based on the asymptotic
MHS solution and the exact MHS solution for a linear force-free
case and two different MHS cases, i.e. we will vary mainly the
parameter $a$.

\subsection{Analytical multipole case} 
\label{sec:AnMuResults}

We consider the analytical boundary condition described in Section \ref{sec:AnMu} and consider seven 
different parameter combinations
of our model, which are listed in
Table \ref{tab:AnMu}. 
For all configurations 
we have used a normalising magnetic 
field strength of $B_0 = 500$ Gauss. 
All calculations are carried out on a grid with
the resolution $n_x = n_y = 200$ and $n_z = 400$.
We include the first $200$ modes in both the $x$- and the
$y$-directions in the Fourier series.
With $b=1.0$, $\tilde{\alpha} = 0.05$, as well as 
$\tilde{k}^2_{min} = (\pi/L_x)^2 + (\pi/L_y)^2 \approx 0.049348$ one obtains from Equation \ref{eq:amax} that the 
maximum value of $a$ is
$a_{max} \approx 0.4746697$. 
Based on this value,
we have chosen test cases with $a = 0.22$ and $a = 0.44$.
This allows us 
to 
investigate how 
the 
amplitude of the perpendicular current density
affects the quality of the approximation of the exact solution
by the asymptotic solution.

%
%
%
We investigate the effect of varying the parameter $\Delta \tilde{z}$ by comparing extrapolation results obtained
using the standard value of $\Delta \tilde{z} = 0.1 \tilde{z}_0$ with results for
$\Delta \tilde{z} = 0.5 \tilde{z}_0$ and $\Delta \tilde{z} = \tilde{z}_0$.
These choices of $\Delta \tilde{z}$ are more extreme cases, purposefully chosen to test the 
limitations
of 
using the asymptotic solution.
%
%
The differences between the exact solution results and the asymptotic solution results are a consequence
of the widening of the domain over which the transition of the exact solution from a non-force-free state
to a force-free state takes place. The asymptotic solution is independent of $\Delta\tilde{z}$ (see 
Equation \ref{eq:DPhiBarAsymp} and
hence remains unchanged if the value of $\Delta\tilde{z}$ changes.
Additionally, for 
$\Delta \tilde{z} = 0.1 \tilde{z}_0$ we have included a linear force-free case ($a=0.0$).
So overall we investigate one linear force-free and six MHS 
cases
in total (Table \ref{tab:AnMu}). 
For each of these 
cases
we compare the results based on the exact solution 
with those based on
the asymptotic solution. 

\begin{table}
\caption{Quantitative comparison between the exact solution by \cite{Neukirch2019} and the 
asymptotic
solution presented in Section \ref{sec:methodology} using the analytical multipole 
described in equation \ref{eq:VonMises} as boundary condition. (1) a linear force free 
configuration; (2-3) MHS configurations using a realistic value of $\Delta z = 0.1z_0$ 
together with the two different values of $a$; (4-5) MHS configurations using the smaller 
value of $a$ together with increased $\Delta z = 0.5z_0$ or $\Delta z = z_0$; (6-7) MHS 
configurations using the larger value of $a$ together with increased $\Delta z = 0.5z_0$ or $\Delta z = z_0$. 
We compare the vector correlation ($C_{Vec}$), the angular difference ($C_{CS}$), 
the complement of the normalised vector error ($1-E_n$), the complement of the mean 
vector error ($1-E_m$), the relative total magnetic energy ($\varepsilon$), the field line divergence ($l_{Div}$) and the Pearson correlation for plasma pressure ($r_P$) and density ($r_D$). 
}

\begin{tabular}{c>{\columncolor{lavender}}cc>{\columncolor{lavender}}cc>{\columncolor{lavender}}cc>{\columncolor{lavender}}c} 
  \hline
 Case & 1 & 2 & 3 & 4 & 5 & 6 & 7  \\
  \hline	
  $a$  & $0.0$ & $0.22$ & $0.44$ & $0.22$ & $0.22$ & $0.44$ & $0.44$ \\
   $\Delta \tilde{z}/\tilde{z}_0$ & $0.1$ & $0.1$ & $0.1$  & $0.5$ & $1.0$ 
   & $0.5$& $1.0$ \\
   \hline
  $C_{Vec}$ & $1.0$ & $1.0$ & ${1.0}$ & 
  $0.9998$ & $0.9993$ & $0.9982$ & $0.9939$ \\
  $C_{CS}$ & ${1.0}$ & ${1.0}$ & ${1.0}$ & 
  ${1.0}$ & $0.9999$ & $0.9999$ & $0.9995$ \\
   $1-E_n$ & ${1.0}$ & $0.9989$ & $0.9965$ & $0.9845$ 
          & $0.968$ & $0.9552$ & $0.9133$ \\
  $1-E_m$ & ${1.0}$ & $0.9996$ & $0.9991$ & $0.9932$ 
        & $0.98$ & $0.9845$ & $0.9568$  \\
  $\varepsilon$ & ${1.0}$ & $1.0002$ & $1.0024$ & 
     $1.004$ & $1.007$ & $1.0346$ & $1.061$  \\
  $l_{Div}$ & ${1.0}$&${1.0}$ &${1.0}$ & 
            $0.9986$ & $0.9986$ & 0.9993 & $0.9986$ \\
  $r_P$  & - & ${1.0}$ & ${1.0}$& ${1.0}$ & 
         $0.9999$ & $0.9998$ & $0.9991$ \\
  $r_D$  & - & ${1.0}$ & ${1.0}$ & ${1.0}$ & 
               ${1.0}$& $0.9999$ & $0.9999$ \\
   \hline
\end{tabular}
\label{tab:AnMu}
\end{table}

Table \ref{tab:AnMu} summarises our results. For the linear force-free 
case (case 1)
the magnetic fields are identical
due to $f_{N+W}(\tilde{z}) = 0$ for all $\tilde{z}$. Hence,
all figures of merit for the magnetic fields have a value of $1.0$.
A comparison of the pressure and density profiles does not make sense
for this case because one would simply be comparing the stratified
background model with itself.


The following two cases (cases 2 and 3) show the 
metrics for $\Delta \tilde{z} = 0.1 \tilde{z}_0$, 
but values of $a=0.22$ (case 2)
and $a=0.44$ (case 3). For these cases, small deviations of less 
than $1\%$ are recorded for the two error metrics $E_n$ and $E_m$, and
the magnetic energy metric $\varepsilon$. Overall, these
figures indicate that the asymptotic MHS solution is an excellent
approximation of the exact MHS solution for these parameter values.


For cases 4 to 7 the values of $\Delta\tilde{z}$ are increased
to $\Delta \tilde{z} = 0.5 \tilde{z}_0$ (cases 4 and 6) and
to $\Delta \tilde{z} = \tilde{z}_0$ (cases 5 and 7).
The values for $a$ are again $0.22$ (cases 4 and 5)
and $0.44$ (cases 6 and 7).
Table \ref{tab:AnMu} shows that for all these cases the error metrics $E_n$ and
$E_m$ deviate most from the ideal value of $1.0$. For $a=0.22$ (cases 4 and 5)
they are the only figures of merit that show a larger than $1\%$ deviation from the ideal
value, although as expected all figures of merit show an increase in the differences
between the exact and the asymptotic solution cases with increasing $\Delta\tilde{z}$. 
When the value of $a$ is doubled to $0.44$, all figures of merit indicate
that the difference between the exact and asymptotic case are generally larger. As before
these differences
increase with increasing value of $\Delta\tilde{z}$. In particular the error metrics
indicate more significant levels of discrepancy between the two solutions for these cases,
although the values of the metrics are still surprisingly close to $1.0$ even for 
case 7 with $\Delta \tilde{z} = \tilde{z}_0$, which very strongly pushes the limit
for the validity of the asymptotic solution.

\begin{figure}
\centering
\includegraphics[width=0.8\textwidth,clip=]{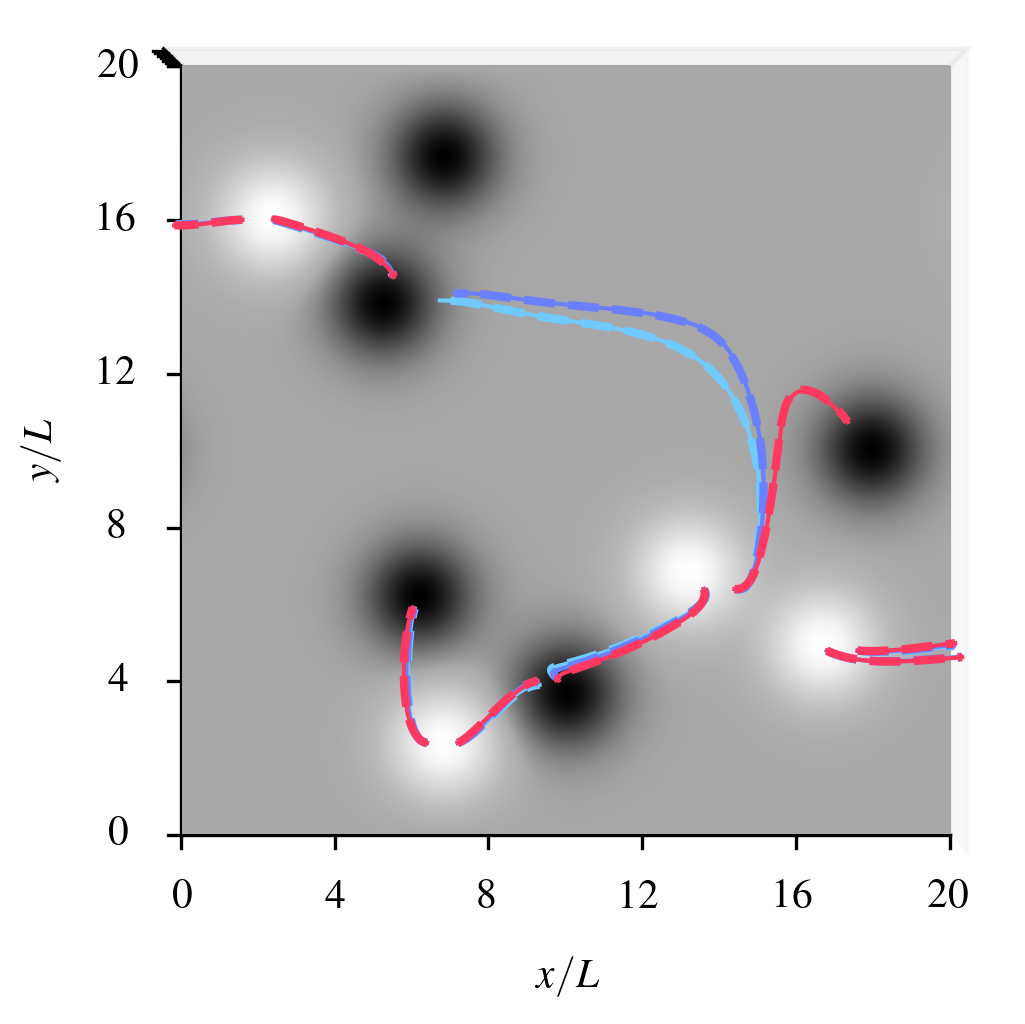}
\includegraphics[width=0.8\textwidth,clip=]{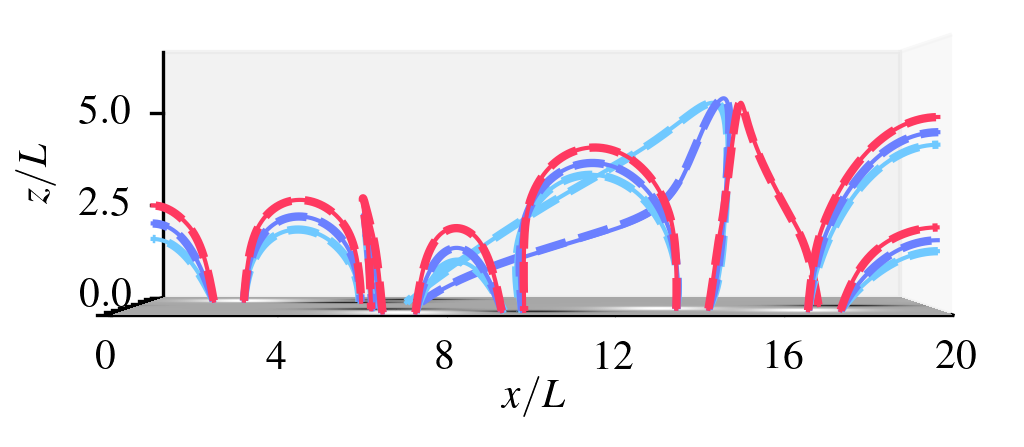}
\caption{Comparison between field lines resulting from a selection of 
footpoints for the exact solution (dashed) 
and the 
asymptotic
solution (solid) for 
cases 1
(light blue), 
2 (dark blue),
and 
3 (red) 
projected onto
the $x$-$y$-plane (top panel, top-down view) and the $x$-$z$-plane (bottom panel). 
If the thick dashes are centred around the thin solid lines the 
solutions match best, whereas a larger distance between the dashed
and solid lines of the same colour indicate a larger difference between
the solutions. The shape of most field lines shown is the same for all three cases.
The only noticeable difference between the three cases is that of the field lines
starting from a footpoint towards the right side of the positive (white) source
region centred at $(\tilde{x}, \tilde{y})\approx (13.2, 6.8)$. For this footpoint
the field line of linear force-free field ends at the negative (black) polarity region
closest to the right hand side of the box, whereas the field lines for all MHS cases
end at a negative polarity region in the top left corner. One can also notices 
in both panels that 
for this footpoint 
the field lines for the two MHS cases differ significantly. 
}
\label{fig:AnMuFieldlines}
\end{figure}

Figure \ref{fig:AnMuFieldlines} shows a comparison 
between 
the exact and the asymptotic solution for 
cases 1, 2, and 3.
For this comparison we display a selection of magnetic field lines
starting from the same footpoints for each of the three cases.
Generally, larger values of $a$ 
lead
to an increase in 
steepness of the
field lines below $\tilde{z}_0$.
Therefore, 
for $a = 0.44$ (case 3)
the magnetic field lines
reach greater heights than the 
linear force-free field (case 1) 
and the MHS field for
$a = 0.22$ (case 2). 
The footpoints of the field lines displayed have been 
chosen to 
emphasise the differences between the 
cases,
especially in height and 
steepness below $\tilde{z}_0$. 
Additionally, for the field lines starting at one of the footpoints 
the
connectivity changes between the 
linear force-free case and the MHS 
cases (see caption of Figure  \ref{fig:AnMuFieldlines}).
We remark that the 
asymptotic
solution 
successfully 
matches
the connectivity of the field line of the exact solution.
For this field line the 
largest
difference between the exact and the 
asymptotic solution 
happens for case 2,
as 
the dashed line appears to be minimally further away from the solid line than in case 3. 

\begin{figure}
\centering
\includegraphics[width=0.49\textwidth,clip=]{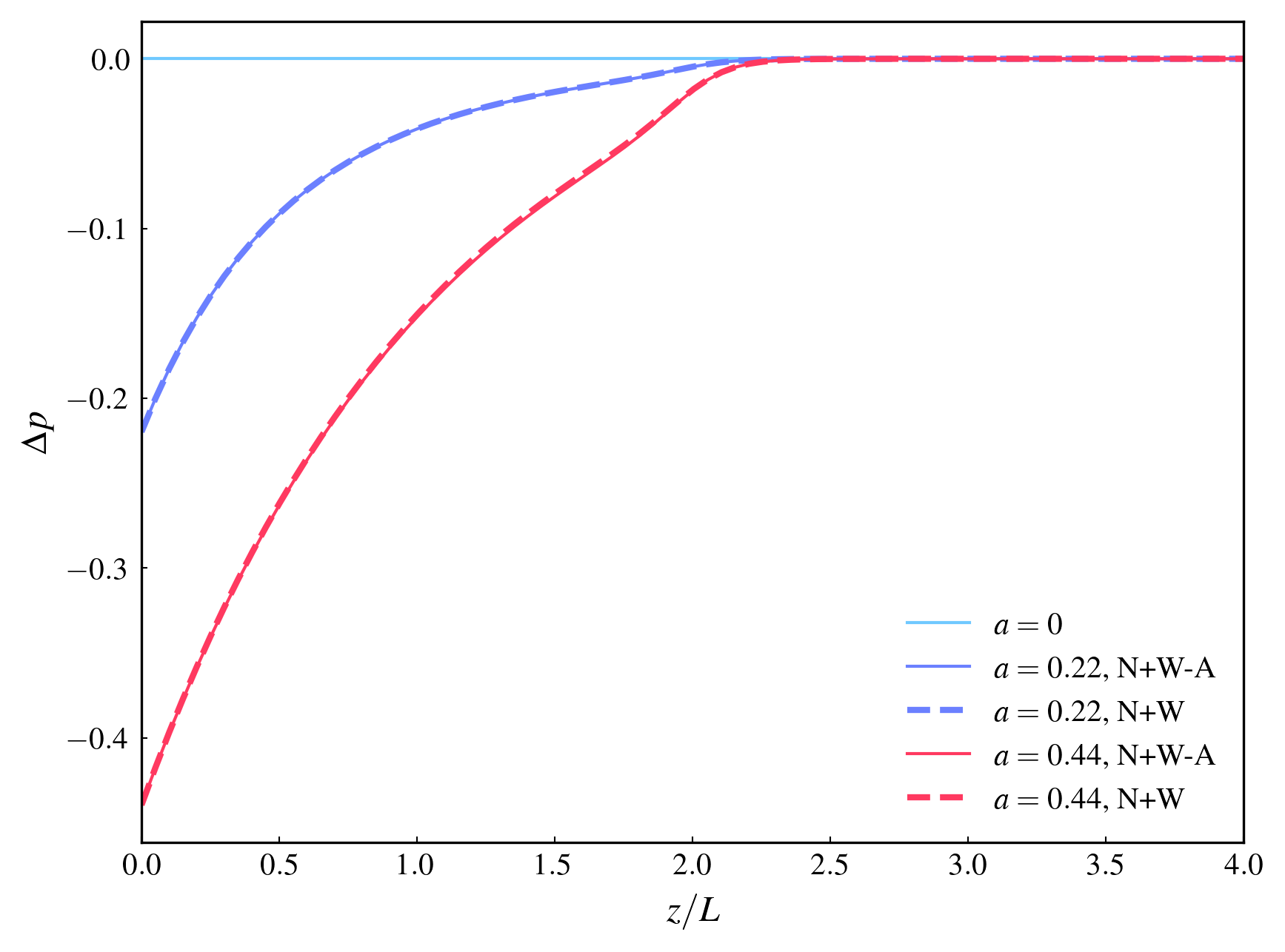}
\includegraphics[width=0.49\textwidth,clip=]{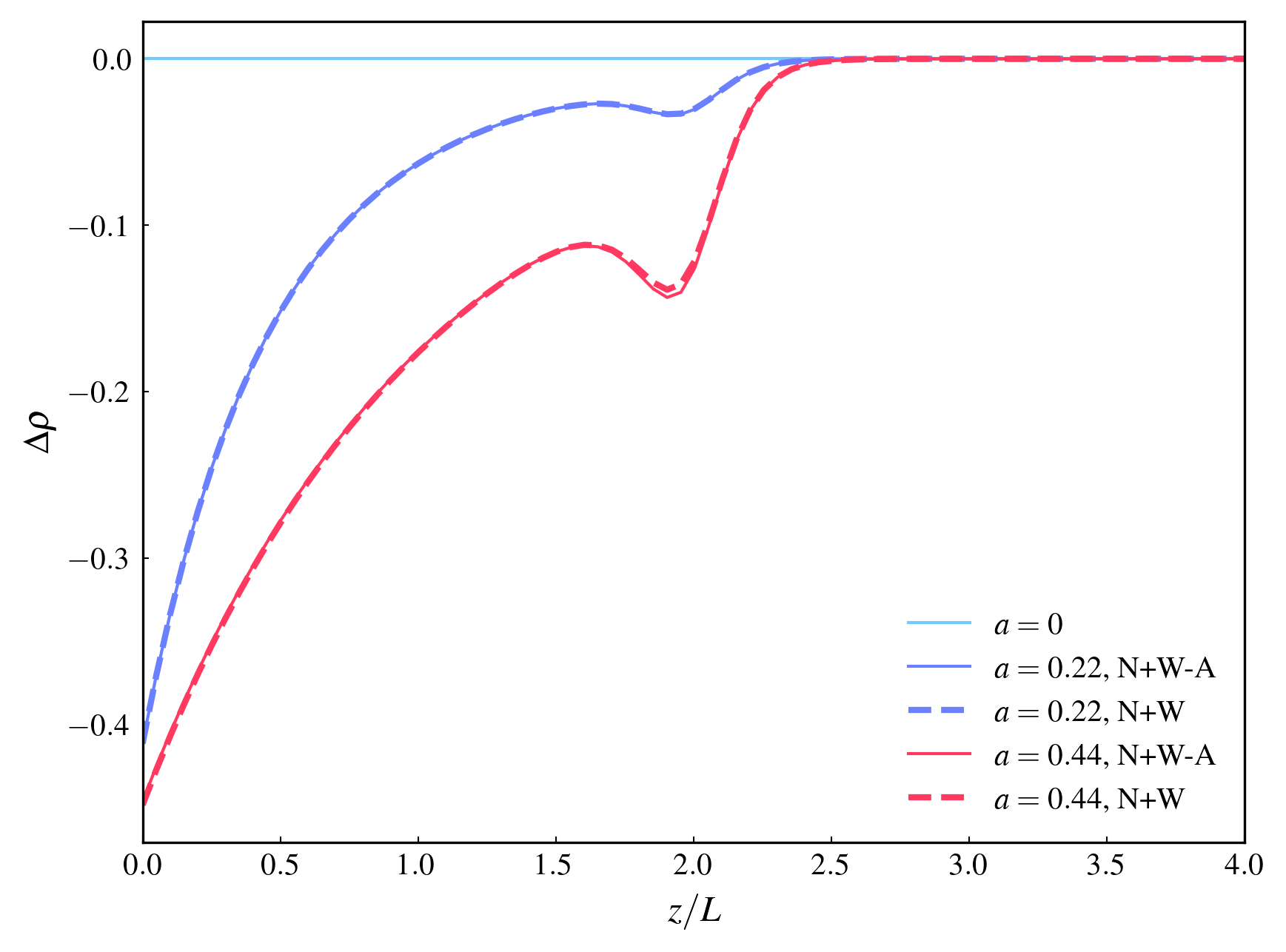}
\caption{Pressure (left) and density (right) variations of 
cases 1
(light blue), 
2
(dark blue) and 
3
(red) for approximate (solid) and exact (dashed) solution at $\tilde{x}$ and $\tilde{y}$ 
where $B_z$ is maximal on the photosphere ($\tilde{z}$=0). $\Delta p$ is 
normalised by $B_{max}^2/\mu_0$ and $\Delta \rho$ normalised 
by $B_{max}^2/(\mu_0 g_S L)$, where $B_{max} = \mathrm{max}(B_z(x,y,0))$.}
\label{fig:AnMuPD}
\end{figure}

Figure \ref{fig:AnMuPD} 
shows 
a comparison between
the exact (N+W) and the asymptotic (N+W-A) solutions
of
the 
variation with height ($\tilde{z}$) of $\Delta p$ and $\Delta \rho$ 
for cases $1$, $2$, and $3$.
The linear force-free solution (case 1) is only
provided
as a reference case, because by definition $\Delta p$ and $\Delta \rho$ both identically vanish for this case.
To make the comparison easier we show the variation of $\Delta p$ and $\Delta \rho$ with height
at fixed values of $\tilde{x}$ and $\tilde{y}$.
These fixed values of $\tilde{x}$ and $\tilde{y}$ have been chosen such that
for this values $|B_z|$ takes on its maximal value at $\tilde{z} =0$.
This position has been chosen, because this is where
we expect $\Delta p$ and $\Delta \rho$ to take on 
their largest absolute values (we note that both generally have 
negative values).
The local 
minimum
below $\tilde{z}_0$ in the density variation is 
caused by the term containing 
the derivative of $f$ in the 
definition of $\Delta \rho$ (see Equation \ref{eq:rho3Dfull}). 
To avoid the negative values of 
$\Delta p$ and $\Delta \rho$ to cause the full
pressure and density becoming negative, care
has to be taken when picking the background
atmosphere \citep[see e.g.][for an example]{Wiegelmann2015}.
We also note from Figure \ref{fig:AnMuPD} that
the largest differences between the exact (N+W)
and asymptotic (N+W-A) solutions 
can be seen
around and closely to the
local minimum of $\Delta p$ and $\Delta \rho$.
The reason for this can be traced back to the 
differences between the exact and the asymptotic
solutions for $\bar\Phi$ and its first derivative $\bar\Phi^\prime$,
which are also largest around $\tilde{z}\approx \tilde{z}_0$, as shown in Figure \ref{fig:phi}.
The slightly larger deviation seen in $\Delta \rho$ compared to $\Delta p$ is due to it having
one term which depends on $\bar\Phi^\prime$, and the difference between the exact solution and the asymptotic
solution is a bit larger for $\bar\Phi^\prime$ than for $\bar\Phi$.

The analysis above has established that for a sensible choice 
of parameters, a magnetic field extrapolation based on the asymptotic
solution (N+W-A) approximates results based on the
exact solution (N+W) very accurately. This leads to the
next question: does the use of the asymptotic solution lead
to savings in runtime which would justify its use in 
comparison with using the exact solution.

The difference in computational effort between the exact and the
asymptotic solutions results from the evaluation of
the function $\bar{\Phi}(\tilde{z})$ and its first derivative,
$\bar{\Phi}^\prime(\tilde{z})$.
Therefore, we test the 
difference in
numerical efficiency between the exact and the asymptotic 
solution by 
evaluating 
$\bar{\Phi}(\tilde{z})$ and $\bar{\Phi}^\prime(\tilde{z})$
on a grid in $\tilde{z}$ of size $n_z = 400$. We also 
need to test the effect on the computing time of
the dependence of 
$\bar{\Phi}(\tilde{z})$ on $\tilde{k}^2$, which we have suppressed
for ease of notation.
This is done
by using 
$n_f=400$ Fourier modes in each of the 
horizontal directions ($\tilde{x}$ and $\tilde{y}$).
In summary, our test grid has the size 
$n_f^2 \cdot n_z = 400^3$.
%
%
For the efficiency comparison we have used the parameter
values $\tilde{z}_0=2.0$, $\Delta \tilde{z} = 0.1\tilde{z}_0$ 
$\alpha=0.05$, $a=0.22$ and $b=1.0$, which corresponds to case $2$
in Table \ref{tab:AnMu}.
The 
numerical calculations 
have
been carried out on a 
MacBook Air (2020) M1 processor with 16 GB RAM. 

Table \ref{tab:NumEff} summarises the 
results of all runtime tests (for more details see Appendix \ref{sec:appendix-a}). These results indicate a
significant advantage in computing time for the 
asymptotic
solution. In this case, the calculation 
of the 
asymptotic
solution was per execution an 
order of magnitude faster than that of the exact solution. 
The combined computation times of function $\bar{\Phi}$
and its derivative are $2.31$ seconds for the 
approximate solution and $41.4$ seconds for the exact solution. 
This implies a time advantage of $94.47\%$ over
our a $400^3$ 
grid and with the current implementation of the numerical code.


\begin{table}
\caption{Computation time comparison of $\bar{\Phi}$ and 
its first derivative
for the exact solution (N+W) and
the asymptotic solution (N+W-A). Shown are the average
time $\pm$ standard deviation (per loop) for 10 runs 
(100 for-loops each), on a $400^3$ grid (MacBook Air M1, 2020). 
The numbers show
that the asymptotic solution has a significant advantage in 
computational efficiency over the exact solution for this test case.
For details of the methodology used see the main text
and Appendix \ref{sec:appendix-a}. }
\label{tab:NumEff}
\begin{tabular}{ccc} 
  \hline
   & N+W & N+W-A \\ 
  \hline
  \rowcolor{lavender}
  $\bar{\Phi}$ & $17.8$ s $\pm 14$ ms & $1.15$ s $\pm 4.76$ ms \\ 
  $d \bar{\Phi} / dz$ & $23.2$ s $\pm 19.5$ ms & $1.16$ s $\pm 1.21$ ms \\
  \hline
\end{tabular}
\end{table}

\subsection{SDO/HMI magnetogram case}
\label{sec:SDOresults}
In this subsection, we will present results of
the very first application of 
our extrapolation method based on
the \citet{Neukirch2019} MHS solution
family to 
boundary conditions
based on observational
data. As in Section \ref{sec:AnMuResults}
we will compare the extrapolation results based on
the exact MHS solution with those based on
the asymptotic solution.
The 
SDO/HMI magnetogram
we shall use has been presented and discussed in Section \ref{sec:Obs}.


For the extrapolation we have included $239$ Fourier modes in both the
$x$- and the $y$-direction. In the $z$-direction our domain extends to
$\tilde{z} = 20.0$. We use a grid with $222$ points in the $z$-direction.

In our choice of parameter values we will be guided by 
the insights gained in Section \ref{sec:AnMuResults}. 
As usual for this paper we use $b=1.0$, and, as before, 
$\tilde{z}_0=2.0$. As shown in Section \ref{sec:AnMuResults},
a value of $\Delta \tilde{z} = 0.1 \tilde{z}_0$ leads 
to a very accurate representation of the exact solution by the asymptotic
solution, and it is also in approximate agreement with the height and
width of the solar transition region. We shall use 
a value of $\tilde{\alpha} =0.01$. 
%
%
We will compare extrapolation results for three
different values of the parameter $a$, namely 
a linear force-free case with $a=0.0$, and two
MHS cases with $a=0.19$ and $a=0.38$, respectively.

\begin{table}
\caption{Quantitative comparison between the exact solution 
and the 
asymptotic
solution 
for
SDO/HMI magnetogram as
boundary condition.
Shown are the values of the figures of merit defined in
Section \ref{sec:analysistools}
for a linear force-free ($a=0.0$) and two MHS configurations. 
All values indicate an excellent
agreement between the exact and the asymptotic solution.
}
\label{tab:SDO}

\begin{tabular}{ccccccccc}
  \hline
  $a$  & $C_{Vec}$ & $C_{CS}$ & $1-E_n$ & $1-E_m$ & $\varepsilon$ & $l_{Div}$ & $r_P$ & $r_D$ 
  \\
  \hline	
  \rowcolor{lavender}
  0.0 & 1.0
  & 1.0
 & 1.0
 & 1.0
 & 1.0
 & 0.9994
 & - & - 
  \\
  0.19 & 
  & 1.0
  & 0.9999 & 0.9999 & 1.0002 
  & 0.9994
  & 1.0
  & 1.0
  \\
  \rowcolor{lavender}
  0.38 & 1.0
  & 1.0
  & 0.9998 & 0.9998 & 1.0001 & 0.9991 
  & 1.0
  & 1.0
  \\
  \hline
\end{tabular}
\end{table} 

Table \ref{tab:SDO} summarises the results of our comparison showing that in this case the exact and the approximate model reach almost identical results in all configurations. 

\begin{figure} 
\centering
    \includegraphics[width=0.8\textwidth]{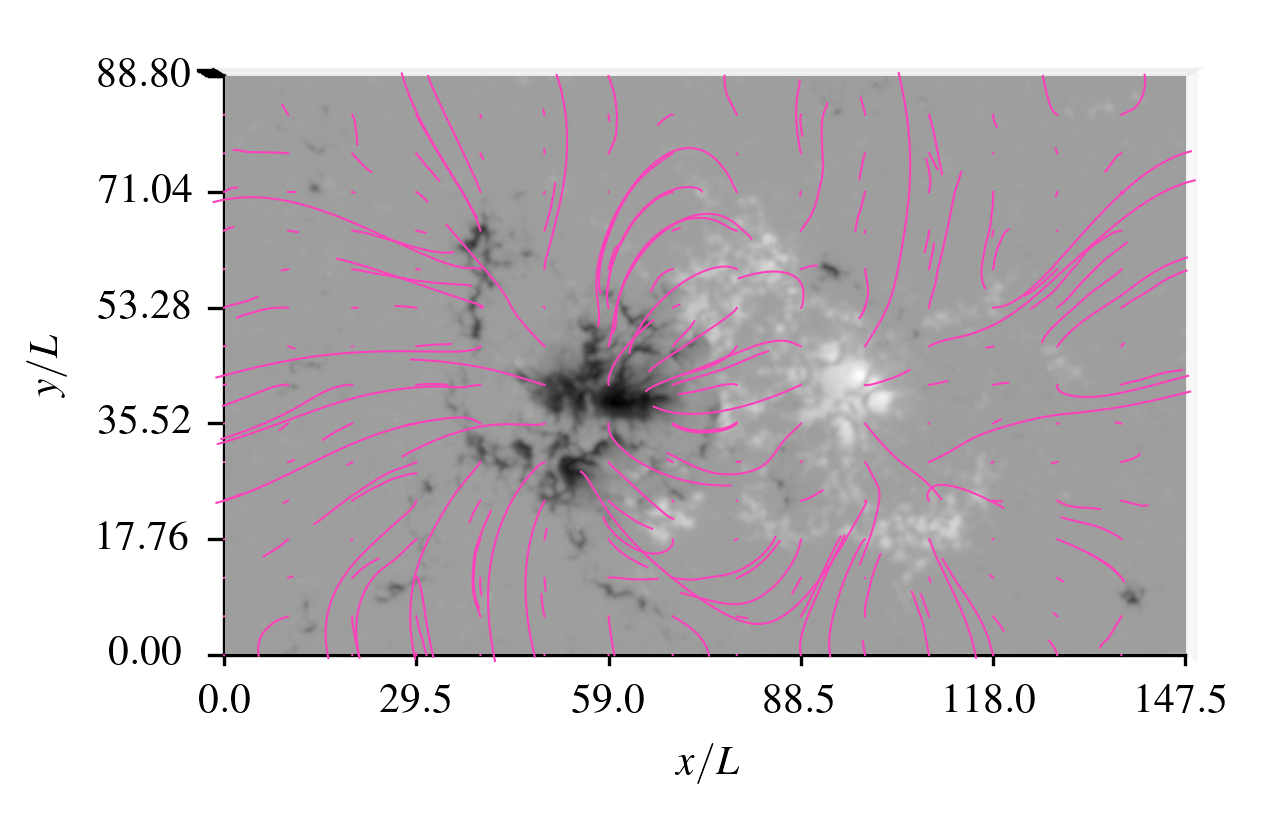}
    \includegraphics[width=0.8\textwidth]{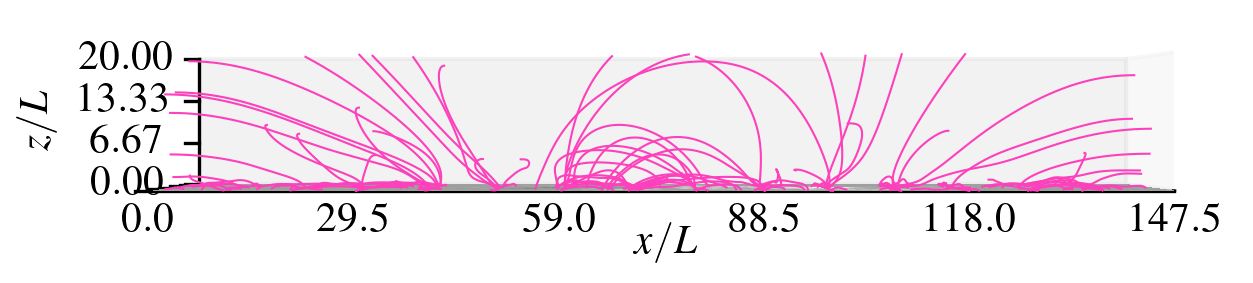}
    \caption{Field lines in the $x$-$y$-plane (top) and the $x$-$z$-plane (bottom) for low-a MHS configuration.}
    \label{fig:SDOFieldlines}
\end{figure}

\begin{figure} 
\centering
    \includegraphics[height=0.48\textwidth]{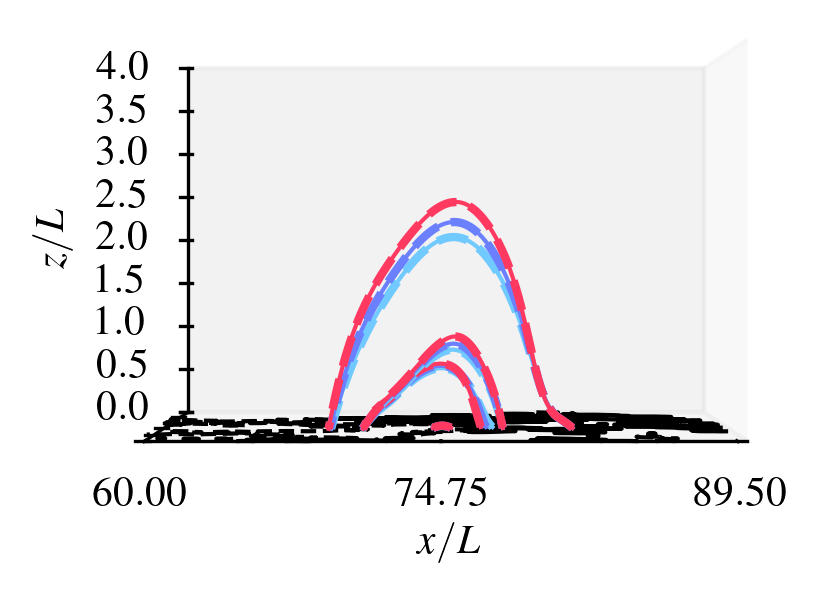}
    \includegraphics[height=0.48\textwidth]{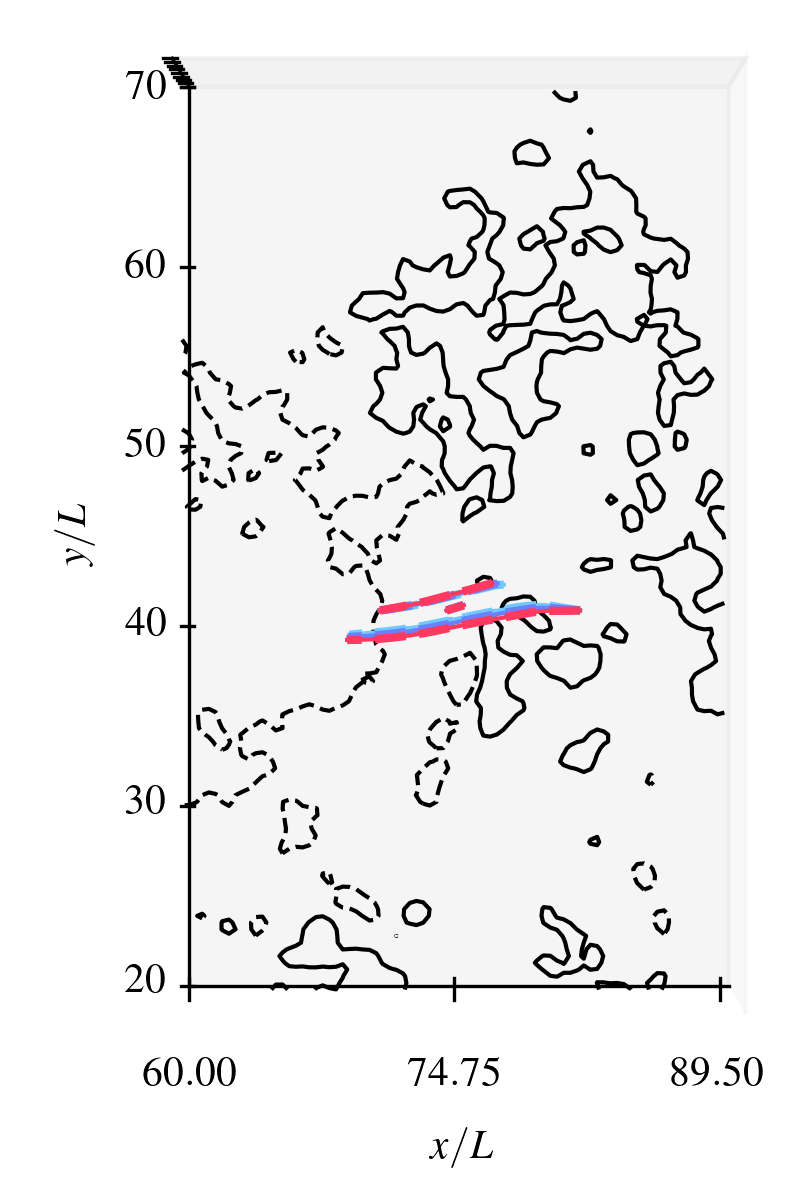}
    \caption{Comparison of magnetic field lines starting 
    at
    four
    selected 
    footpoints in the $\tilde{x}-\tilde{y}$-plane ($\tilde{z} =0$).
    The
    field lines are shown from a perspective
    looking in the direction of the $\tilde{y}$-axis 
    (left panel) and looking down
    the $\tilde{z}$-axis
    (right panel).
    The field lines for the linear force-free case ($a=0$) are
    shown in light blue, for the $a=0.19$ MHS case in dark blue,
    and for the $a=0.39$ MHS case in red. Dashed lines represent field lines resulting from utilisation of the exact solution, solid lines for utilisation of asymptotic approximation.}
    \label{fig:SDOFieldlinesZoom}
\end{figure}

For reference, Figure \ref{fig:SDOFieldlines} shows 
the 
magnetic field structure
obtained from the
extrapolation using the asymptotic solution
for the 
$a=0.19$
MHS configuration from a top-down perspective and a side-view 
perspective along the $\tilde{y}$-axis. 
Figure \ref{fig:SDOFieldlinesZoom} shows field lines 
at 
four selected footpoints
for the three different values of $a$. 
The differences between 
the field lines for different values
of $a$ is most obvious in the left panel, 
particularly for the field lines reaching somewhat larger heights
(note that $\tilde{z}_0 = 2.0$ is the midpoint of
the $\tilde{z}$-axis in the panel on the left).
These differences are caused by
steepening of field lines 
below $\tilde{z}_0$
with increasing $a$.

 \begin{figure} 
 \centering
     \includegraphics[width=0.49\textwidth]{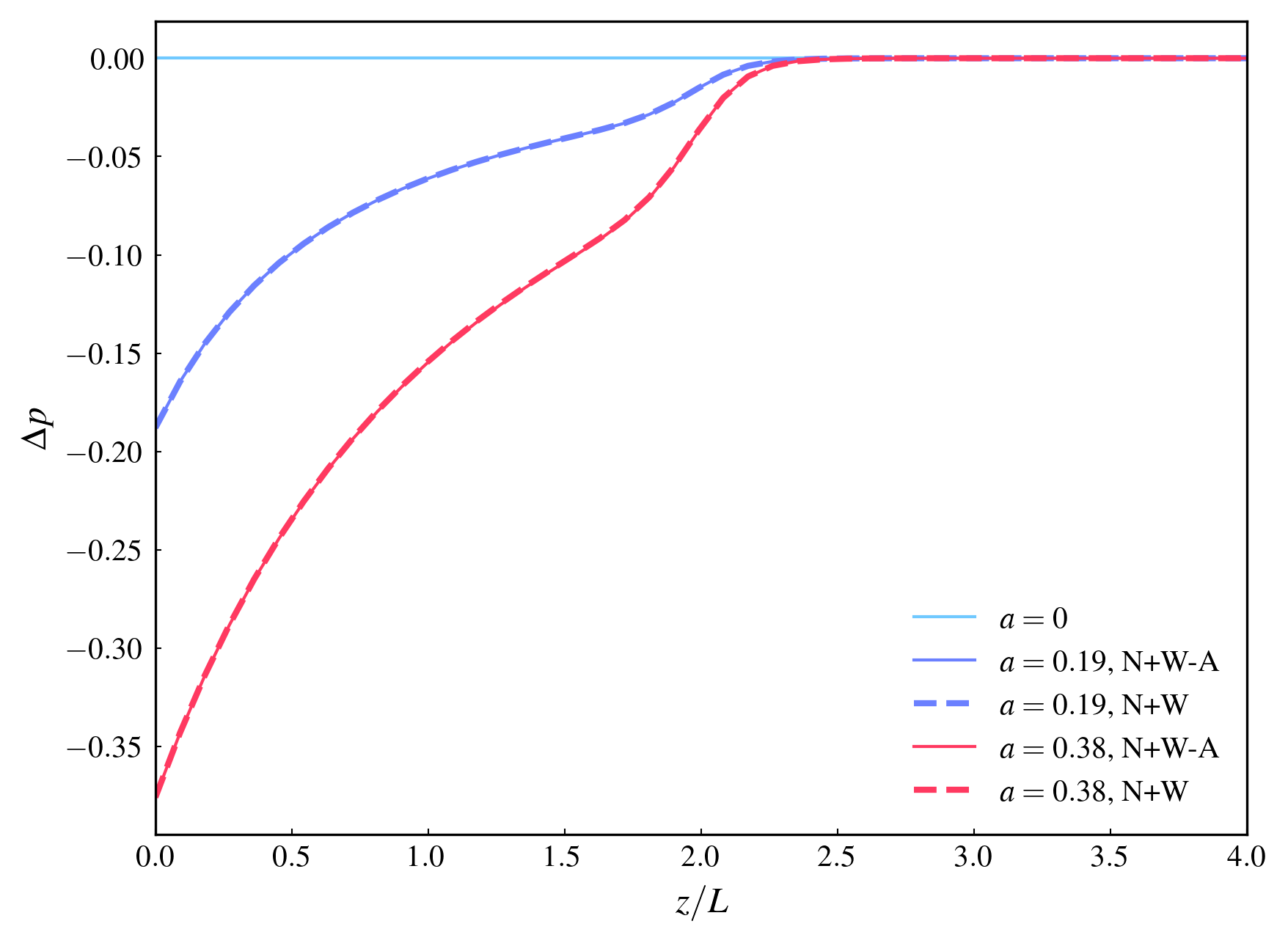}
     \includegraphics[width=0.49\textwidth]{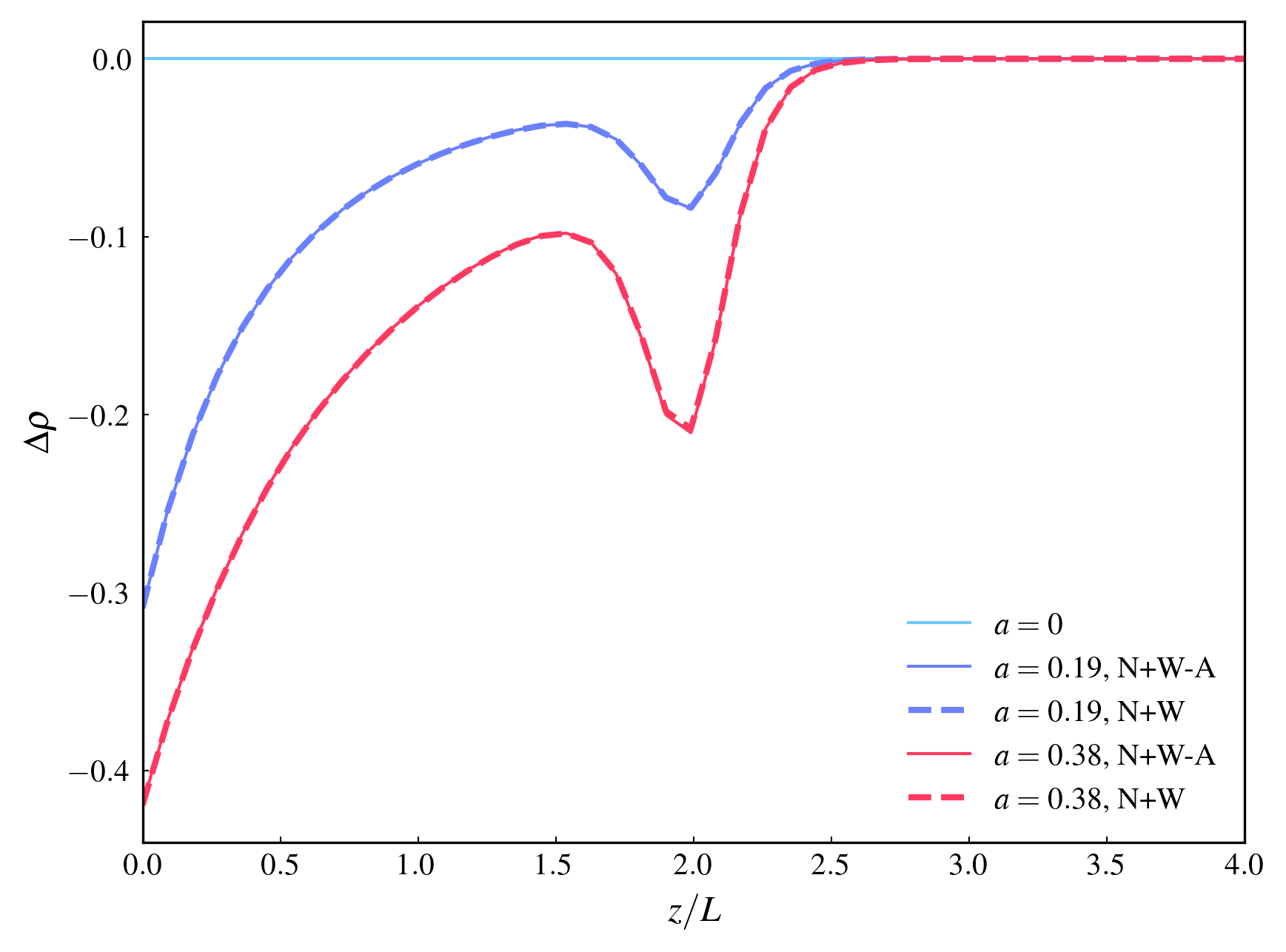}
     \caption{Pressure (left panel) and density (right panel) 
     variations for the linear force-free (light blue) 
     and two MHS configurations ($a=0.19$, dark blue), 
     $a=0.38$, red) at $\tilde{x}$ and $\tilde{y}$ where $|B_z|$ is 
     maximal on the photosphere. $\Delta p$ is 
     normalised by $B_{max}^2/\mu_0$ and 
     $\Delta \rho$ normalised 
     by $B_{max}^2/(\mu_0 g L)$, where $B_{max} = \mathrm{max}(|B_z|, 
     z=0)$. Dashed lines correspond to the exact solution, 
     solid lines to the asymptotic solution.} 
     \label{fig:SDOPD}
 \end{figure}
In figure \ref{fig:SDOPD} we see the spatial variation 
of pressure and density with height $\tilde{z}$ at $\tilde{x}$ 
and $\tilde{y}$ 
where $| B_z |$ is maximal on the photosphere. 
Both pressure and density variation display behaviour 
that is very similar to
the 
example based on
the analytical multipole 
boundary conditions. 
Overall, all the results based on the SDO/HMI magnetogram data
corroborate the conclusion that the asymptotic
MHS solution is an excellent approximation of the exact MHS solution
for reasonably chosen parameter values

\begin{table}
\caption{Comparison of the computation time of 
magnetic field vector $\textbf{B}$ and 
the partial derivatives of $B_z$ for
the exact MHS solution (N+W) and the asymptotic MHS
solution (N+W-A). The results show a clear computational
advantage is gained by using the asymptotic solution.}
\label{tab:SDO-runtime}
\begin{tabular}{lll}
  \hline
  $a$  & N+W & N+W-A\\
  \hline
  \rowcolor{lavender}
  0.0 & 3m 23.0s & 38.6s\\
  0.19 &  13m 25.8s & 1m 26.3s\\
 \rowcolor{lavender}
  0.38 &  13m 31.3s & 3m 39.8s\\
  \hline
\end{tabular}
\end{table} 
It remains to investigate whether the gain in computational
time of using the asymptotic solution compared to the exact solution
can also be found for the extrapolations based on 
our observational magnetogram. We have applied the same methodology
that we used for the analytical multipole boundary condition.
Table \ref{tab:SDO-runtime} 
shows the computational time needed for the exact (N+W)
and the asymptotic (N+W-A) solutions for all 
three configurations for the SDO/HMI boundary conditions.
The results clearly confirm
that 
the asymptotic 
solution
provides a significant runtime advantage over
the exact solution also in this case. 

\section{Summary and Conclusion}
\label{sec:summary}

In this paper we have presented a way to significantly improve the efficiency of MHS magnetic field extrapolation based on
the family of analytical MHS solutions by \citet[][]{Neukirch2019}. These analytical MHS solutions allow
for a transition between a non-force-free and a force-free domain in the direction of height above the photosphere. We have 
achieved this improvement in efficiency by finding a new and simple asymptotic solution which approximates the
exact solution very well as long as the width of the transition zone between the non-force-free and the force-free domain is
sufficiently small. The quality of the approximation of the exact solution by the asymptotic solution has been
assessed by comparing extrapolation results for two different boundary conditions: one artificial magnetogram based on
a period multipole structure, and an SDO/HMI magnetogram with a dominant magnetic bipole structure. For both cases, we
carried out parameter studies and found using figures of merit and visual comparisons of field lines, pressure and density plots
that
the results based on the asymptotic solution approximate the results using the exact solution very well if
the width and height of the non-force-free to force-free transitional zone are chosen to be 
in the same range as those
of the transition region of the solar atmosphere.

After establishing that using the asymptotic solution does not lead to incorrect results, we have tested the
gain in computational speed by using it instead of the exact solution. For both magnetogram case and all tested
parameter combinations we found that the speed-up is significant (a factor $10$ or more).

Applying our extrapolation code to the SDO/HMI data is actually the very first time that the \citet{Neukirch2019} MHS
solutions have been used for extrapolation based on observational data. We are in the process to create a Python library
for magnetic field extrapolation
based on the work presented in this paper, which we plan to make publicly available. 

We emphasise that 
extrapolation methods based on analytical, linear MHS solutions such as the one by \citet{Neukirch2019}
should not be 
regarded
as a 
replacement
for numerical nonlinear MHS extrapolation methods. 
When using extrapolation methods based on analytical MHS solutions one has to bear in mind 
that they are restricted by the number of 
assumptions necessary to obtain them
and by the unavoidable introduction of free parameters. On the other hand, they are 
computationally
much more efficient and can therefore be useful as a "quick-look" tool, which is complementary
to the more computationally heavy nonlinear MHS extrapolation methods.

\begin{acks}
The authors thank ISSI/ISSI-BJ for support via International Team 568 "Magnetohydrostatic 
Modeling of the Solar Atmosphere with New Datasets" and the team's participants for useful discussions.
LN acknowledges financial support by the School of Mathematics and Statistics, University of St Andrews,
throughout her PhD. TN acknowledges financial support by the UK's Science and Technology
Facilities Council (STFC) via Consolidated Grants ST/S000402/1 and ST/W001195/1. The authors thank the anonymous reviewer for helpful and constructive comments. 
\end{acks}

\appendix   

\section{Details of computational efficiency comparison}
\label{sec:appendix-a}

We have used the IPython magic shell \texttt{\%timeit} to time 
the execution time of the calculation. We have 
set \texttt{\%timeit} to average the 
results over 10 repeats of 100 loops each. 
One loop in \texttt{\%timeit} covers the 
computation of the function values in the whole 
discretised and truncated Fourier space. 
Each such loop in \texttt{\%timeit} includes a 
for-loop over the $z$-axis of length 400, 
in which for each $z$, the function values are 
calculated simultaneously for all $k_n$ and $k_m$ 
through vectorisation. Hence, each loop in the 
for-loop (not in \texttt{\%timeit}!) corresponds to 
calculating the function values for a horizontal 
plane $(k_n, k_m)$ in the Fourier space 
at a specific $z_{i_z}$. Therefore, 10 runs of 100 
loops correspond to overall 1000 calculations of 
the whole space per function. 

The parameters $\delta$, $\gamma$, 
$\tilde\delta$ and $\tilde\gamma$, as seen in 
the definitions of exact and asymptotic versions of 
$\bar\Phi$ 
(Equations \ref{eq:PhiBarNW} and \ref{eq:PhiBarAsymp}, 
respectively),
are calculated independently beforehand 
for each Fourier mode.
They are then 
passed directly to the 
functions whose evaluation is to be timed, 
such that the 
comparison
exclusively measures the calculation 
time of $\bar\Phi$ and $\bar\Phi^\prime$ for
each Fourier mode.

\begin{authorcontribution}
LN wrote the extrapolation code and did the numerical computations. Both authors contributed to the
theoretical calculations, the analysis of the results and the writing of the paper.
\end{authorcontribution}

\begin{fundinginformation}
This work was supported by the UK's Science and Technology
Facilities Council (STFC) via Consolidated Grants ST/S000402/1 and ST/W001195/1.
\end{fundinginformation}

\begin{dataavailability}
The data used in this paper are publicly available at 
\linebreak
http://jsoc.stanford.edu/.
\end{dataavailability}



\begin{ethics}
\begin{conflict}
The authors declare no competing interests.
\end{conflict}
\end{ethics}

\bibliographystyle{spr-mp-sola}
\bibliography{sorted.bib}  

\end{document}